\documentclass[fleqn,usenatbib]{mnras}
\usepackage{newtxtext,newtxmath}

\usepackage[T1]{fontenc}
\usepackage{longtable}

\usepackage{graphicx}
\usepackage{bm}
\usepackage{natbib}
\usepackage{booktabs}
\usepackage{color}
\usepackage{units}
\usepackage{orcidlink}
\usepackage{hyperref}
\usepackage{xcolor}

\usepackage[normalem]{ulem}

\newcommand{\Msun}{M_{\odot}}
\newcommand{\Mni}{M_{\rm Ni}}
\newcommand{\Mej}{M_{\rm ej}}
\newcommand{\Ek}{E_{\rm k}}
\newcommand{\fmix}{f_{\rm mix}}

\newcommand{\tdiff}{t_{\rm diff}}

\newcommand{\kappagamma}{\kappa_{\gamma}}
\newcommand{\fdep}{f_{\rm dep}}
\newcommand{\snmix}{\program{snmix}}
\newcommand{\jax}{\program{JAX}}
\newcommand{\redback}{\program{Redback}}
\newcommand{\redbackjax}{\program{Redback-JAX}}
\newcommand{\program}[1]{\textsc{#1}}
\newcommand{\citeg}[1]{\citep[e.g.,][]{#1}}
\newcommand{\be}{\begin{equation}}
\newcommand{\ee}{\end{equation}}

\title[Radioactive structure in supernovae]{Signatures of $^{56}$Ni Mixing and Neutron-rich Ejecta in Supernovae}

\author[N.~Sarin]{Nikhil Sarin$^{1, 2}$\thanks{E-mail:nsarin.astro@gmail.com}\orcidlink{0000-0003-2700-1030}, 
\\
$^{1}$Kavli Institute for Cosmology, University of Cambridge, Madingley Road, CB3 0HA, UK\\
$^{2}$Institute of Astronomy, University of Cambridge, Madingley Road, CB3 0HA, UK\\}

\date{Accepted XXX. Received YYY; in original form ZZZ}

\pubyear{\the\year{}}

\begin{document}
\label{firstpage}
\pagerange{\pageref{firstpage}--\pageref{lastpage}}
\maketitle

\begin{abstract}
Supernova lightcurves are often interpreted with one-zone radioactive-decay models that ignore a key variable that can affect interpretation and inferred parameters: the distribution of radioactive material. Using a multi-shell model, we explore the impact of $^{56}$Ni mixing in supernovae and r-process material in collapsars. Moving $^{56}$Ni outward reduces the overlying diffusion column, producing faster and brighter rises at fixed $\Mni$, $\Mej$, and $\Ek$, and changes the tail through local gamma-ray leakage. A fast, bright rise is not, by itself, evidence for low ejecta mass or a requirement for engine power, with significant overlap between highly mixed and engine-powered lightcurves. One-zone fits to mixed bolometric light curves produce visually good fits but biased parameters. At fixed opacity, outward mixing is absorbed mainly by low inferred $\Mej$ and high inferred $f_{\rm Ni}$, while $\Mni$ remains stable. If opacity is free, the fully mixed case is recovered with $\kappa^{\rm fit}/\kappa^{\rm input}\simeq0.24$. These shifts affect inferred explosion energies and progenitor mappings, and amplified in photometric fits.
Exploring collapsar r-process enrichment, we find that the signature is not always a NIR excess and depends sensitively on the nickel-powered background, radial placement, angular distribution, and viewing angle of neutron-rich ejecta. In our setup, spherical models often show optical suppression and delayed colour evolution.
Our disk-wind models suggest that fast-rising on-axis GRB-SNe are poor r-process targets for equatorially confined neutron-rich winds, and become constraining only if the r-process material reaches latitudes $\gtrsim 30^\circ$ from the equatorial plane.
\end{abstract}


\section{Introduction} \label{sec:intro}
Light curves of supernovae are often a primary route to inferring physical parameters and therefore the properties of the progenitor. Their rise times, peak luminosities, and post-peak declines are used to infer nickel masses, ejecta masses, and explosion energies. Light-curve fits to superluminous supernovae, broad-lined Ic supernovae (Ic-BL SNe), and gamma-ray burst supernovae (GRB-SNe) are also often used to argue for or against additional central-engine or interaction power~\citep{Cano2017, Gal-Yam2019}. 
Most of these inferences rely, explicitly or implicitly, on one-zone radioactive models. A common starting point is the model of \citet{Arnett1980, Arnett1982}, in which the ejecta are represented by a single effective diffusion timescale and the peak luminosity is tied to the instantaneous radioactive heating rate. These models have been applied to large samples of core-collapse~\citep[e.g.,][]{Drout2011, Taddia2019, Barbarino2021, Srinivasaragavan2024}, and thermonuclear SNe~\citep[e.g.,][]{Scalzo2014, Sarin2026_ia}, and inferences drawn from such fits are then used to compare explosion mechanisms, nucleosynthetic yields, and the relation between different classes of supernovae.

However, one-zone models are simplified. They remove a physical variable that explosions are expected to have and that should vary between explosions and between supernova classes: the spatial distribution of radioactive material. This variable can have significant impact on many different observables in supernovae, including their lightcurves as well as spectra~\citep{Yoon2019, Moriya2020_mixing}. In supernova explosion simulations, $^{56}$Ni is synthesized deep in the ejecta and then mixed outward by Rayleigh--Taylor instabilities, neutrino-driven convection, explosion asymmetries, and jet-driven turbulence~\citep[e.g.,][]{Woosley2002, Hammer2010, Muller2012, Reichert2023}. While the bulk of $^{56}$Ni may remain deep in the ejecta, a non-negligible fraction may propagate to large radii with the fastest-moving ejecta. Different degrees of mixing place the same radioactive power beneath different overlying ejecta masses, effectively lowering the mass the energy has to diffuse through. Centrally concentrated nickel loses more energy to expansion before photons escape, whereas nickel mixed to larger radii can power a faster and brighter peak for the same total nickel mass, $\Mni$. Inferences drawn from one-zone models may then provide incorrect interpretations of parameters without producing obviously bad fits, absorbing the missing radioactive structure into the fitted ejecta mass, nickel fraction, opacity, or velocity.

We first explore how the light curves change when the nickel distribution is varied at fixed $\Mni$, ejecta mass $\Mej$, and kinetic energy $\Ek$, and how one-zone or Arnett-like analyses respond when this degree of freedom is absent in light-curve inference. These inferred parameters are used to compare supernovae to explosion and nucleosynthesis models, and to decide when radioactive heating is insufficient. Fast, luminous Ic-BL and GRB-SNe are a particularly important case as their early lightcurves can depart significantly from one-zone model predictions~\citep[e.g.,][]{Niblett2025}. In one-zone models, a short diffusion time can be interpreted as low ejecta mass, high $\Ek/\Mej$, or used as an argument for an additional power source beyond radioactivity. Nickel mixed to large radii offers another possible route to a fast optical rise, so the radioactive-structure alternative should be tested before the optical light curve alone is used to argue for sustained engine power. Einstein Probe~\citep{Yuan2015} has begun to routinely reveal fast X-ray transients associated with Ic-BL SNe~\citep{Sun2025, vanDalen2025, Rastinejad+25, Rastinejad+26}, including events interpreted with engine-powered models~\citep{Srinivasaragavan2025, Zhu2025}. Although some associated supernovae appear otherwise fairly standard, their light curves already stretch one-zone models into uncomfortable parts of parameter space. The start of new surveys such as LSST~\citep{Ivezic2019}, combined with higher-cadence observations from LS4~\citep{Miller2025} and ZTF~\citep{Bellm2019}, will only increase the sample.

The same ``where is the radioactive material?'' question also matters for r-process material in collapsars. If collapsars synthesize r-process material in neutron-rich disk winds~\citeg{Siegel2019, Gottlieb2023}, that material adds radioactive heating and, if lanthanides are present, high-opacity ejecta that can reshape the optical-NIR SED. \citet{Barnes2022} showed that a spherical calculation with r-process-rich material beneath an r-process-free supernova envelope can produce a detectable NIR excess. We explore how the observable prediction changes if disk-wind material is less deeply buried, confined to the equatorial plane, or mixed to higher latitudes.

Hydrodynamical simulations and radiation-transfer calculations are the right tools for detailed predictions, but they are not practical for the parameter exploration needed here. Inference can require up to $100,000$ model evaluations for one supernova, and this problem adds several poorly constrained dimensions, including nickel placement, opacity, ejecta mass, r-process mass, and viewing angle. Surrogate models or GPU-accelerated radiation transport may eventually make this feasible~\citep{Sarin2025, Sarin2026_csm} but they require a more careful construction for the full problem. Here we use a fast semi-analytic model to isolate these variables and map their effect on the observables.

To study these questions we introduce \snmix, a semi-analytic multi-shell model that captures radial heating and opacity structure while remaining fast enough for inference through JAX~\citep{jax2018github}. The model discretizes homologously expanding ejecta in velocity, evolves shell thermal energies with shell-dependent heating and opacity, computes a multi-component SED, and produces filter-integrated magnitudes. The framework is broader than GRB-SNe, but they are the main application here because they connect nickel mixing, fast optical rises, and collapsar r-process signatures. This paper is structured as follows. In Section~\ref{sec:model} we describe the core nickel-powered model. Section~\ref{sec:ni_mixing} explores nickel mixing and its consequences for one-zone inference, including an analysis of SN~1998bw. In Section~\ref{sec:rprocess}, we introduce the additional r-process ingredients and apply the framework to collapsar r-process signatures. We discuss implications for existing analyses and observing strategies in Section~\ref{sec:discussion} and conclude in Sec.~\ref{sec:conclusions}.
\section{Core Nickel-powered Ejecta Model} \label{sec:model}
In this section we describe the core nickel-powered model. The additional r-process heating, opacity, optically thin SED, and disk-wind geometry are introduced separately in Section~\ref{sec:rprocess_model}, immediately before the r-process results.

\subsection{Ejecta and Radioactive Structure} \label{sec:ejecta}

Our first model assumption is the distribution of ejecta and radioactive nickel. As is common in semi-analytic models, we adopt a broken power-law density profile following \citet{Matzner1999}, parametrized by the ejecta mass $\Mej$, kinetic energy $\Ek$, and two power-law indices $\delta$ (inner) and $n$ (outer):
\begin{equation} \label{eq:density}
    \rho(v, t) = A t^{-3}
    \begin{cases}
        \left(\dfrac{v}{v_t}\right)^{-\delta} & v < v_t \\[10pt]
        \left(\dfrac{v}{v_t}\right)^{-n} & v \geq v_t
    \end{cases}
\end{equation}
where $A$ is set by the requirement that the density integrates to $\Mej$. The transition velocity is
\begin{equation} \label{eq:bpl_transition_velocity}
    v_t =
    \left[
    \frac{2(5-\delta)(n-5)\Ek}
         {(3-\delta)(n-3)\Mej}
    \right]^{1/2}.
\end{equation}
This choice fixes the integrated kinetic energy to $\Ek$ for homologous expansion, $R_i=v_i t$. We adopt $\delta = 1$ and $n = 10$ as representative values for stripped-envelope supernovae~\citep{Matzner1999}. The ejecta are discretized into $N$ shells with geometrically spaced velocities $v_i$ spanning from $v_{\rm min}$ to $v_{\rm max}$ following this density profile. Each shell $i$ has mass $\Delta m_i$, radius $R_i(t) = v_i t$, and width $\Delta R_i(t) = \Delta v_i \, t$.

The $^{56}$Ni mass fraction in each shell is specified by a tophat in enclosed mass coordinate, parametrized by the mixing fraction $\fmix \in [0, 1]$:
\begin{equation} \label{eq:ni_profile}
    X_{\rm Ni}(m) =
    \begin{cases}
        \Mni / (\fmix \Mej) & m \leq \fmix \Mej, \\
        0                  & m > \fmix \Mej,
    \end{cases}
\end{equation}
where $m$ is the enclosed ejecta mass measured outward from the centre. Here $\fmix$ is an extent in mass coordinate. The total nickel mass is fixed to $\Mni = f_{\rm Ni}\Mej$, so the local nickel abundance inside the mixed region is $X_{\rm Ni}=f_{\rm Ni}/\fmix$. Therefore $\fmix = 1$ means the nickel is uniformly distributed through the whole ejecta with $X_{\rm Ni}=f_{\rm Ni}$, not that the ejecta is made entirely of nickel. Small $\fmix$ concentrates the same nickel mass in the innermost mass shells. Models with $\fmix < f_{\rm Ni}$ are excluded because they would require $X_{\rm Ni} > 1$ in the mixed region.

We adopt the top-hat model as our baseline controlled parameterisation because it has few degrees of freedom. Realistic explosions can have smoother cores, plumes, tails, and asymmetric nickel at large radii. As a simple non-top-hat diagnostic, we also use a core-plus-tail profile motivated by two-component models~\citep{Nagy2016},
\begin{equation} \label{eq:ni_core_tail_profile}
    X_{\rm Ni}^{\rm ct}(m) =
    \begin{cases}
        (1-\eta_{\rm tail})\Mni / m_{\rm core}
            & m \leq m_{\rm core}, \\
        \eta_{\rm tail}\Mni / (m_{\rm out}-m_{\rm core})
            & m_{\rm core} < m \leq m_{\rm out}, \\
        0   & m > m_{\rm out},
    \end{cases}
\end{equation}
where $\eta_{\rm tail}$ is the fraction of the total nickel mass placed in the tail, $m_{\rm core}$ is the core boundary, and $m_{\rm out}$ is the outer edge of the tail. As for the top-hat profile, combinations that require $X_{\rm Ni}>1$ are excluded. This profile is included primarily to test whether the conclusions depend on the discontinuous top-hat choice, not to represent a unique explosion calculation.

\label{sec:heating}
The next ingredient is the nickel heating rate. The radioactive decay chain $^{56}$Ni $\to$ $^{56}$Co $\to$ $^{56}$Fe provides the dominant heating source for supernovae, ignoring engines or CSM interaction. The specific heating rate per unit nickel mass is
\begin{equation} \label{eq:ni_heating}
    \dot{q}_{\rm Ni}(t) = \left(\epsilon_{\rm Ni} - \epsilon_{\rm Co}\right) e^{-t/\tau_{\rm Ni}} + \epsilon_{\rm Co}\, e^{-t/\tau_{\rm Co}}
\end{equation}
where $\tau_{\rm Ni} = 8.77\,{\rm d}$ and $\tau_{\rm Co} = 111.3\,{\rm d}$ are the e-folding lifetimes, and $\epsilon_{\rm Ni} = 3.90 \times 10^{10}\,{\rm erg\,s^{-1}\,g^{-1}}$ and $\epsilon_{\rm Co} = 6.78 \times 10^{9}\,{\rm erg\,s^{-1}\,g^{-1}}$ are the total specific energy release rates. 

\begin{figure*}
    \centering   \includegraphics[width=1.00\textwidth]{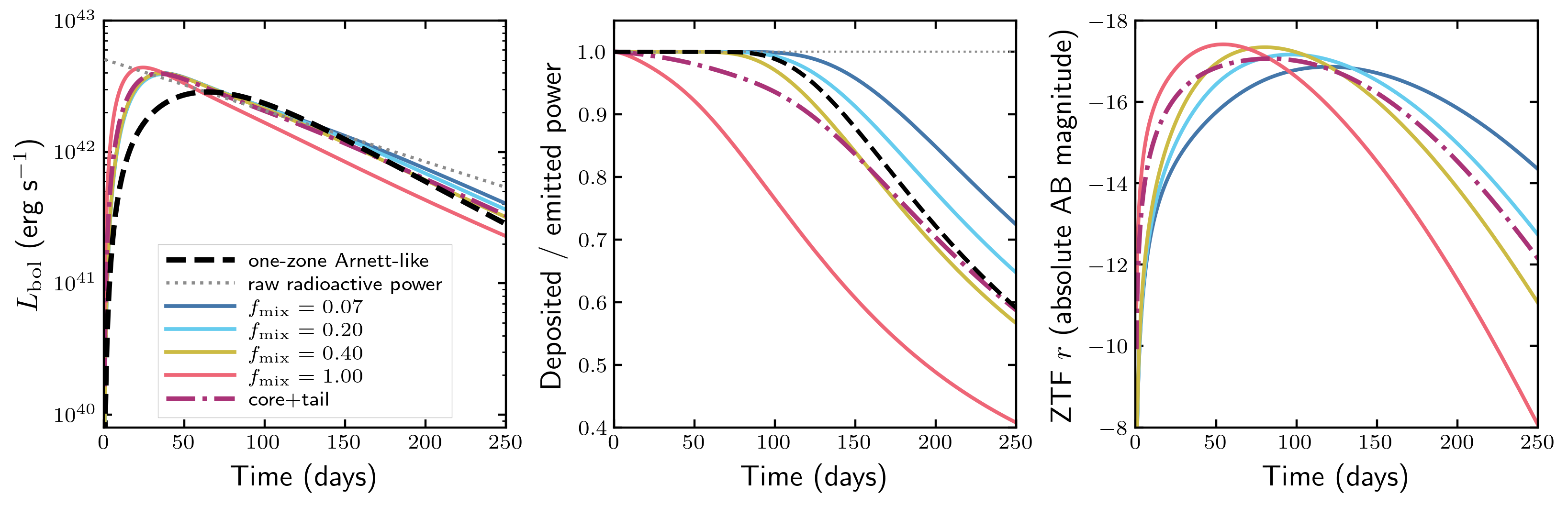}
    \caption{A representative supernova with $\Mej = 5\,\Msun$, $E_{\rm SN} = 1\,{\rm foe}$, $f_{\rm Ni} = 0.07$, and $\kappa = 0.2\,{\rm cm^2\,g^{-1}}$. The mixing coordinate $\fmix$ is the enclosed mass fraction over which the fixed nickel mass is distributed. Solid coloured curves are mass-coordinate top-hats. The dash-dotted curve is a core-plus-tail profile with 80 per cent of the nickel in the inner 20 per cent of the ejecta mass and 20 per cent spread to the surface. The bolometric curves show that outward mixing shortens the rise at fixed ejecta mass and nickel mass. The deposition curves show the corresponding change in radioactive trapping: outward nickel sits under a smaller gamma-ray column and leaks earlier, while minimum-mixing model remains better trapped. The ZTF $r$-band panel contains only the \snmix\ models to isolate the effect of nickel placement. The \redback\ Arnett model uses a global photosphere and temperature with a single characteristic velocity as opposed to the multi-shell SED assumed in \snmix, i.e., differences in photometry between the two models are a combination of multiple effects.}
    \label{fig:mixing_lc_family}
\end{figure*}

\subsection{Transport, Opacity, and Photometry} \label{sec:transport}

We evolve the initial distribution by assuming each shell evolves independently, accounting for radioactive heating, radiative losses, and adiabatic expansion:
\begin{equation} \label{eq:energy}
    \frac{dE_i}{dt} = Q_{{\rm dep},i}(t) - \frac{E_i}{t} - L_i(t)
\end{equation}
where $E_i$ is the thermal energy of shell $i$, $Q_{{\rm dep},i}$ is the deposited radioactive heating rate, $E_i/t$ is the adiabatic loss term for radiation-dominated ejecta in homologous expansion, and $L_i$ is the radiative luminosity escaping from shell $i$. The deposited heating rate is related to the raw heating rate via, 
\begin{equation}
Q_{{\rm dep},i}=\Delta m_{{\rm Ni},i}
\dot q_{\rm Ni}f_{\rm dep},i
+\Delta m_{{\rm rp},i}\dot q_{\rm rp}f_{{\rm dep},i}^{\rm rp}.
\end{equation}
Here $\dot{q}_{\rm Ni}$ is the energy produced via the radioactive decay of nickel and subsequent species, with $f_{{\rm dep},i}$ describing a time-dependent fraction which determines how much of this raw energy is coupled to the ejecta. The second term describes the energy from r-process nucleosynthesis, which we will elaborate on in Sec.~\ref{sec:rprocess_model} and is set to zero in the nickel-only models. The diffusion luminosity is
\begin{equation} \label{eq:diffusion_L}
    L_i = \frac{E_i}{{\tdiff}_i}, \quad {\tdiff}_i = \frac{3\,\tau_i\, R_i}{\beta\, c}
\end{equation}
with $\tau_i = \sum_{j \geq i}\Delta\tau_j$ the cumulative optical depth from shell $i$ to the surface, $\Delta\tau_j = \kappa_j \rho_j \Delta R_j$, $R_i = v_i t$ the shell radius, $c$ the speed of light, and $\beta=13.8$ is the diffusion-geometry constant. The total bolometric luminosity is
\begin{equation} \label{eq:lbol}
    L_{\rm bol} = \sum_i L_i.
\end{equation}
The cumulative optical depth sets the overlying column in the leakage time. This shell-by-shell cumulative-column treatment is the key departure from a one-zone Arnett model and is the source of the systematic differences we explore in Section~\ref{sec:ni_mixing}. Our framework thus far is effectively an extension to one-zone models~\citep{Kasen2010}, similar to the multi-shell toy kilonova model discussed in~\citet{Metzger2020}.

\label{sec:opacity}
The next key ingredient is the opacity. For most of our analysis we adopt a gray, i.e., temperature and wavelength independent, opacity $\kappa = 0.07$--$0.34\,{\rm cm^2\,g^{-1}}$, typical of values used for SN light-curve modelling. This is a standard limitation of semi-analytic models. 

As an extension to this simple implementation, we also include a temperature-dependent opacity law, motivated by numerical simulations~\citep{Nagy2018}, applied shell by shell to the local shell temperature,
\begin{equation} \label{eq:redback_opacity}
    \kappa_i(T_i) =
    \kappa_{\rm min} + \frac{\kappa_{\rm max}-\kappa_{\rm min}}{2}
    \left[1+\tanh\left(\frac{T_i-T_{\rm floor}}{\Delta T}\right)\right],
\end{equation}
with default $\kappa_{\rm min}=10^{-3}\,{\rm cm^2\,g^{-1}}$, $T_{\rm floor}=1000\,{\rm K}$, and $\Delta T=5\times10^4\,{\rm K}$. This is a smooth effective-opacity prescription, not a detailed recombination or ionisation calculation. It lowers the optical diffusion opacity as the ejecta cool, and because it is applied shell by shell it can interact with the nickel distribution. This opacity parameterisation was used in the modelling of SN~2025kg~\citep{Rastinejad+25}.

\label{sec:thermalisation}
Not all energy from radioactive decay will be absorbed by the ejecta. Gamma rays from $^{56}$Ni and $^{56}$Co decay are deposited according to the cumulative column density exterior to the shell where they are produced:
\begin{equation} \label{eq:gamma_dep}
    {\fdep}^{\gamma}_i = 1 - \exp[-3\kappagamma\Sigma_i(t)],
\end{equation}
where $\Sigma_i(t) = \sum_{j \geq i} \Delta m_j/(4\pi R_j^2)$ is the column above shell $i$ and $\kappagamma = 0.027\,{\rm cm^2\,g^{-1}}$~\citep{Colgate1980}. This is the thin-shell approximation to $\int_i^\infty\rho\,dr$ on the homologous grid; using $\Delta m_j/(4\pi R_j^2)$ is equivalent to $\rho_j\Delta R_j$ to leading order for the discretized shells. 
Positrons from $^{56}$Co decay are fully trapped. This prescription is simple, but captures a key distinction from one-zone models. Optical diffusion and gamma-ray deposition are both controlled by shell-dependent properties rather than a single global velocity. Outward nickel therefore tends to reduce the optical diffusion time and the gamma-ray trapping column at the same time; those two effects push the rise and the radioactive tail in different directions.

\label{sec:sed}
To calculate an SED, we assign each shell luminosity a normalized blackbody spectral shape:
\begin{equation} \label{eq:multi_bb}
    L_\lambda = \sum_i L_i\,\phi_\lambda(T_i), \quad
    \int \phi_\lambda(T_i)\,d\lambda = 1,
\end{equation}
where $L_i$ is the shell luminosity from Equation~(\ref{eq:diffusion_L}) and $\phi_\lambda$ is normalized to conserve the bolometric luminosity of that shell. The shell temperature follows from the thermal energy density as $T_i = (E_i / a_{\rm rad} V_i)^{1/4}$. Broadband magnitudes are computed by integrating $L_\lambda$ against standard filter transmission functions via \redback~\citep{Sarin2024}, converting to flux density at the specified luminosity distance, and then converting to bandpass magnitudes.

\label{sec:implementation}
We implement this model, \snmix, as a plugin package for \redbackjax, the \jax-native extension of \redback~\citep{Sarin2024}. The \jax{} framework makes the model differentiable and GPU/CPU agnostic, which is useful for inference studies and large parameter exploration at low computational cost. We also implement a simpler \program{numpy}-only alternative in \redback{}; this version has comparable CPU run time but cannot be run on a GPU. 
\section{Nickel Mixing in Supernovae} \label{sec:ni_mixing}
We now isolate the effect of nickel placement. Unless stated otherwise, $\Mni$, $\Mej$, $\Ek$, and opacity are fixed, so changes in the light curve come only from moving the same radioactive mass through the ejecta.

As stressed above, $\fmix$ is the enclosed ejecta mass fraction reached by nickel, not the nickel abundance itself. For the top-hat profile, $f_{\rm Ni}=0.07$ with $\fmix=1$ gives $X_{\rm Ni}=0.07$ throughout the ejecta, while $\fmix=0.2$ gives $X_{\rm Ni}=0.35$ in the inner 20 per cent by mass and zero outside. In real supernovae, we expect neutrino-driven convection and Rayleigh--Taylor instabilities to carry $^{56}$Ni outward, while weaker mixing or different asymmetries may leave it centrally concentrated~\citep[e.g.,][]{Muller2012, Reichert2023}. Engine activity can also inflate central bubbles and lead to nickel mixing to high mass fractions~\citep{Eiden2026}. We do not model the hydrodynamic origin of $\fmix$ here but simply explore the impact on light curves. We note that one may expect the nickel distribution to be tied to the explosion energy or explosion mechanism~\citep{Woosley2002}, which we also ignore here. 

\subsection{How mixing shapes light curves} \label{sec:mixing_lcs}

In general, the rise time of a supernova scales as
\begin{equation}
    t_{\rm rise} \sim t_{\rm diff,Ni} \simeq
    \left(\frac{\kappa M_{>{\rm Ni}}}{\beta c v_{\rm Ni}}\right)^{1/2},
    \label{eq:mixing_diffusion_scaling}
\end{equation}
where $v_{\rm Ni}$ is the mass-weighted velocity of the nickel-bearing shells. This is the usual diffusion scaling, but with the total ejecta mass replaced by $M_{>{\rm Ni}}$, which describes the total ejecta mass ahead of the heating source, i.e., ahead of the nickel mass. Mixing decreases $M_{>{\rm Ni}}$ and increases $v_{\rm Ni}$, so the same $\Mej$ and $\Ek$ can produce a much shorter rise. As the heating source has to overcome less adiabatic loss, the peak energy is also brighter.

\begin{figure}
    \centering
    \includegraphics[width=\columnwidth]{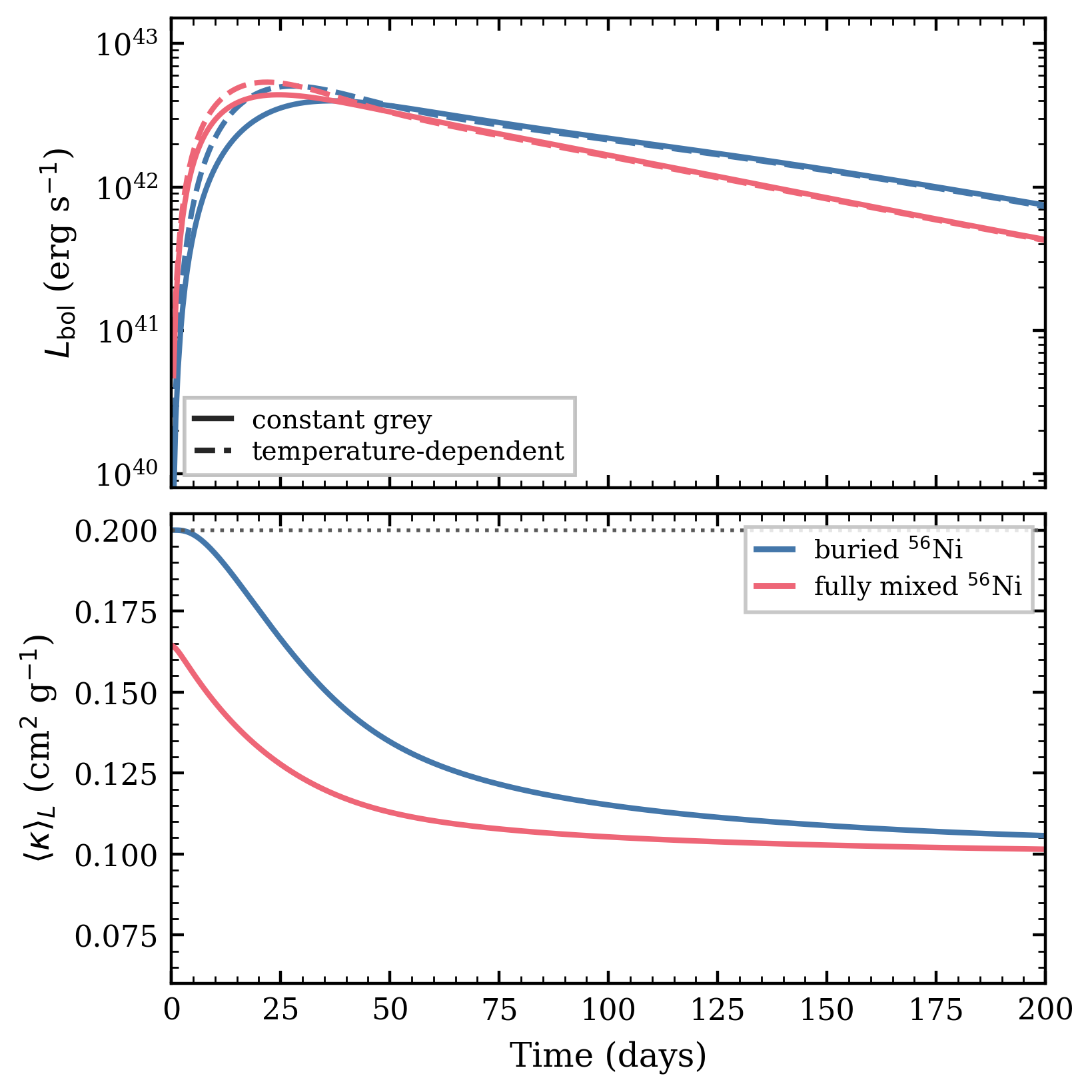}
    \caption{Effect of the temperature-dependent SN opacity law at fixed $\Mej=5\,\Msun$, $E_{\rm SN}=1\,{\rm foe}$, $f_{\rm Ni}=0.07$, and $\kappagamma=0.027\,{\rm cm^2\,g^{-1}}$. Colours distinguish minimum-mixing ($\fmix=f_{\rm Ni}=0.07$) model from a fully mixed nickel ($\fmix=1$) model. Solid curves use constant grey opacity; dashed curves use Equation~(\ref{eq:redback_opacity}). The bottom panel shows the luminosity-weighted opacity in the temperature-dependent run. Lower effective opacity produces faster and brighter peaks at fixed nickel placement.}
    \label{fig:opacity_law}
\end{figure}

\begin{figure*}
    \centering
    \includegraphics[width=0.95\textwidth]{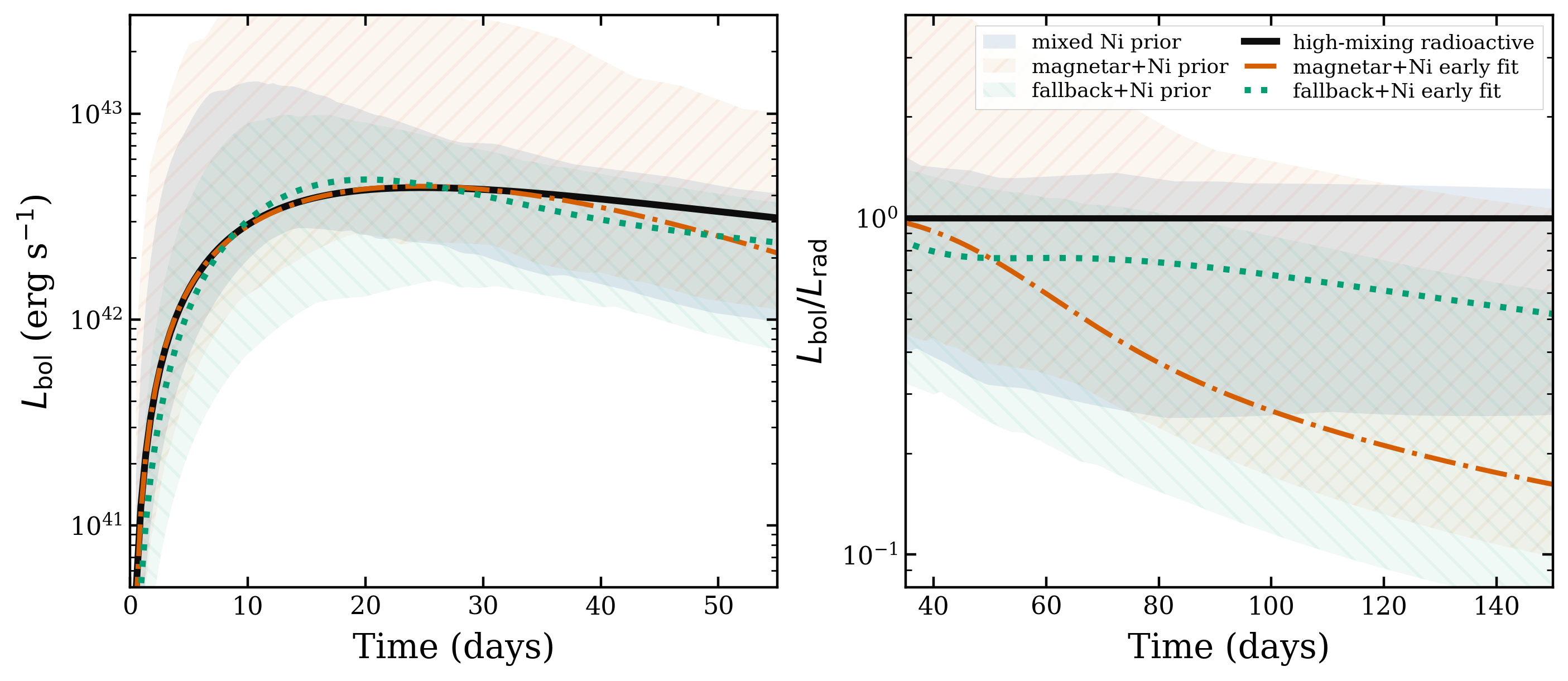}
    \caption{Bolometric degeneracy between outward nickel mixing and simple engine-powered templates. The left panel shows prior-predictive intervals for mixed radioactive, magnetar+Ni, and fallback+Ni models, plus one high-mixing radioactive example and engine fits matched over 1-35 d. The right panel shows late-time luminosity relative to the high-mixing radioactive curve. Early peaks overlap; late-time coverage helps test whether the event follows radioactive heating with gamma-ray leakage or needs sustained engine power.}
    \label{fig:mixing_vs_engine}
\end{figure*}

This basic behaviour is shown in Fig.~\ref{fig:mixing_lc_family}. All curves are for the same SN with $\Mej=5\,\Msun$, $E_{\rm SN}=1\,{\rm foe}$, $f_{\rm Ni}=0.07$, and $\kappa=0.2\,{\rm cm^2\,g^{-1}}$, with different $\fmix$. Centrally concentrated nickel rises slowly because the photons diffuse through a large overlying column. As $\fmix$ increases, more heating occurs at lower column density and the rise becomes faster. The core-plus-tail profile, with $m_{\rm core}=0.2\Mej$, $m_{\rm out}=\Mej$, and $\eta_{\rm tail}=0.2$, falls between the top-hat cases. This comparison highlights that the light curve depends on the full overlying column above the radioactive material, not only on the outermost radius where nickel is present. The black curves show the widely used one-zone Arnett model, which is markedly different to \snmix\, especially at early times but closer at late times (for low values of mixing). This is to be expected as even at the minimum allowed value, $\fmix=f_{\rm Ni}$, \snmix\ retains separate shell reservoirs and cumulative optical and gamma-ray columns, whereas the Arnett model uses one reservoir, one characteristic velocity, and global diffusion and trapping scales.

The second panel illustrates the behaviour of the radioactive tail. A minimum-mixing model, has a heating source that remains under a larger gamma-ray column leading to a more gradual decline, while outward nickel leaks earlier. In the representative model the deposition fraction at 180 d falls from $\simeq0.90$ for a minimum-mixing model to $\simeq0.53$ for fully mixed model, compared with $\simeq0.78$ for the one-zone Arnett calculation. Outward mixing can therefore make the peak faster and brighter while making the radioactive tail fainter. The right panel shows \snmix\ ZTF/r band light curves. These absolute magnitude light curves follow the same general trend as the bolometric lightcurves, where higher mixing leads to brighter and faster peaking supernovae. The core-plus-tail case also shows that bolometric and band-limited light curves need not rank the profiles in the same order, because the nickel profile changes shell temperatures as well as total luminosity. In particular, the core + tail peak is higher in the bolometric lightcurve than all top-hat profiles with $f_{\rm mix} < 1$, while in $r$ band it is dimmer than all curves except $f_{\rm mix}<0.2$. 

These results immediately show that a fast rise is not exclusively a signature of low ejecta mass. In one-zone interpretations, this is often used as evidence for low $\Mej$, but it may instead indicate a high degree of mixing, especially when combined with a fast drop away from the raw radioactive-decay heating rate at late times. 

The above curves highlight the impact of nickel mixing with a fixed gray opacity. However, the opacity is expected to evolve significantly in time and to vary with wavelength. In particular, effective opacities are expected to evolve with ionisation, composition, line blanketing, and the temperature. Since high degrees of mixing reduce the density above the heating source, lowering $\kappa$ reduces the diffusion time for radioactive shells. This can make opacity a strong secondary ingredient for highly mixed sources. 

In Figure~\ref{fig:opacity_law} we show the impact on the lightcurves of the temperature-dependent opacity law in Equation~(\ref{eq:redback_opacity}). We keep $\Mej$, $\Ek$, $f_{\rm Ni}$, and $\kappagamma$ fixed, and compare constant-opacity and temperature-dependent runs for a minimum-mixing and fully-mixed model. The luminosity-weighted opacity in the bottom panel is a diagnostic
\begin{equation}
    \langle\kappa\rangle_L =
    \frac{\sum_i \kappa_i L_i}{\sum_i L_i},
\end{equation}
which summarizes the luminosity-weighted effective opacity of the emitting shells as a function of time. 
For the representative parameters, this drops below the fixed $\kappa=0.2\,{\rm cm^2\,g^{-1}}$ value within the first ten days. As a result, the bolometric peak is brighter and occurs faster. 
The minimum-mixing model moves from $t_{\rm peak}=36.7$ d and $L_{\rm peak}=4.0\times10^{42}\,{\rm erg\,s^{-1}}$ to 27.7 d and $5.0\times10^{42}\,{\rm erg\,s^{-1}}$. The fully mixed model moves from 24.6 d and $4.4\times10^{42}\,{\rm erg\,s^{-1}}$ to 22.0 d and $5.4\times10^{42}\,{\rm erg\,s^{-1}}$. The temperature-dependent law therefore reduces the peak-time separation between these two mixing limits but does not remove the effect of nickel placement. Real opacity evolution is expected to be more complex and wavelength dependent, but this temperature-dependent profile highlights how the opacity and nickel mass distribution can couple to create even faster rises and brighter lightcurves. 
\subsection{Mixing as an alternative to central engines} \label{sec:mixing_vs_engines}
The recent discoveries of Ic-BL SNe enabled by Einstein Probe are broadening the parameter space for GRB-SNe, and several of these supernovae have already been interpreted with magnetar engines~\citep{Sun2025, Srinivasaragavan2025, Zhu2025}. More broadly, engine-based models are a classic crutch in scenarios where one-zone radioactive-decay models fail to explain the early fast rise or brighter than expected luminosity. Common engine-based models include magnetar models~\citep{Kasen2010, Nicholl2017, Sarin2022magnetar, Omand2024}, and fallback-accretion onto a black hole~\citep{Dexter2013, Moriya2019}. In general, more caution should be applied to these interpretations: the optical light curve alone, especially at early times, is a weak discriminator between models~\citep{Moriya2026}. 
Additional evidence at late times or from X-ray and/or radio is critical to determining the nature of the engine.  

We highlight this degeneracy in Fig.~\ref{fig:mixing_vs_engine}, where we show how a well-mixed nickel model can match the rise time and peak luminosity of common parameterisations of engine-powered models~\citep{Sarin2024}. In a one-zone radioactive fit, the same rise forces low $\Mej$ or high $f_{\rm Ni}$, which may lead to dismissal of the model as a plausible solution and be taken as evidence for an engine-powered event. As discussed above, mixing offers another route to brighter peaks and faster rise times by shortening the diffusion column without lowering the total ejecta mass or requiring an uncomfortable fraction of nickel.

In Fig.~\ref{fig:mixing_vs_engine} we also draw prior-predictive bolometric light curves for mixed radioactive models, magnetar+Ni templates, and fallback+Ni templates over broad SESN-like ranges, then fit representative engine templates to a high-mixing radioactive curve over 1-35 d. The early-time regions overlap considerably, demonstrating the degeneracy between these lightcurves. Notably, the separation between these lightcurves appears later, where radioactive models follow the $^{56}$Co budget and gamma-ray leakage, while engines can retain spin-down or fallback power with their own characteristic evolution. This makes the combination of early and late-time data critical, as either phase alone has significant overlap. Late optical photometry can test whether the luminosity follows the radioactive power budget and expected gamma-ray leakage, but it does not always uniquely identify the physical origin of an excess. Persistent or rebrightening X-ray emission, or radio emission requiring ongoing energy injection or interaction, would favour an engine or CSM interaction over a purely mixed radioactive model. Conversely, weak X-ray/radio emission does not rule out an engine, but it makes the optical-only case for sustained power less compelling.

\begin{figure*}
    \centering
    \includegraphics[width=0.95\textwidth]{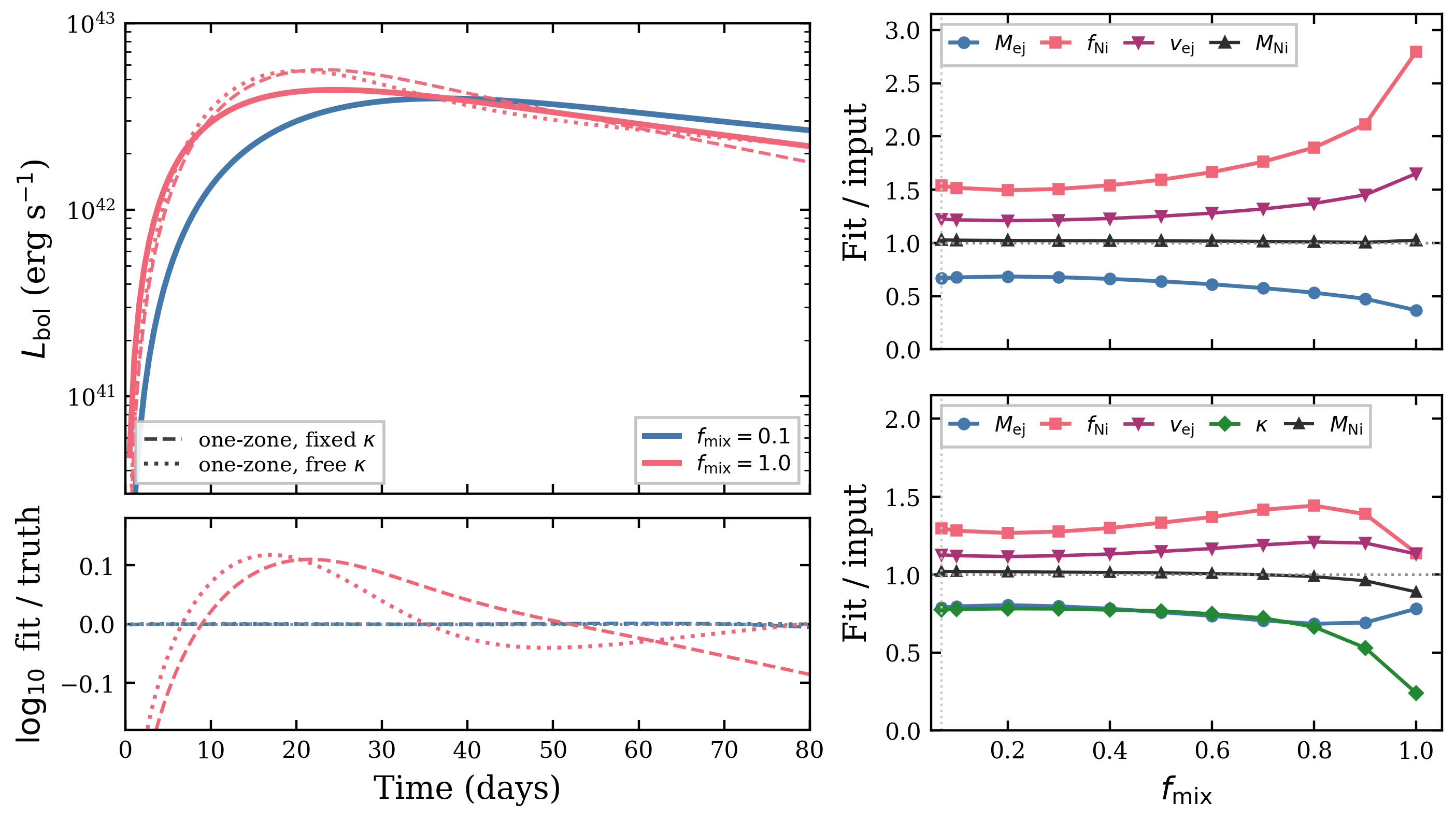}
    \caption{Misspecified one-zone recovery of mixed-ejecta bolometric light curves. The left panels show two synthetic \snmix\ light curves and their best internal one-zone fits; dashed curves fix $\kappa=0.2\,{\rm cm^2\,g^{-1}}$, while dotted curves fit $\kappa$, with the lightcurve fit residuals in the bottom left panel. With realistic uncertainties, these residuals may not by themselves identify the model misspecification. The upper-right panel shows one fitting strategy, where $\kappa$ is fixed and the fitted $\Mej$, $f_{\rm Ni}$, $v_{\rm ej}$, and $\Mni$ are divided by the true synthetic inputs. The lower-right panel shows a second fitting strategy, where $\kappa$ is also fitted. The dotted horizontal line marks exact recovery, and the vertical dotted line marks the minimum physical top-hat value, $\fmix=f_{\rm Ni}$. At fixed opacity, outward mixing is traded mainly against lower $\Mej$ and higher $f_{\rm Ni}$ while $\Mni$ remains comparatively stable. When opacity is free, part of the mismatch moves into lower effective $\kappa$.}
    \label{fig:ni_bias}
\end{figure*}
\subsection{Bias in inference with one-zone models} \label{sec:ni_bias}

\begin{figure*}
    \centering
    \includegraphics[width=1.00\textwidth]{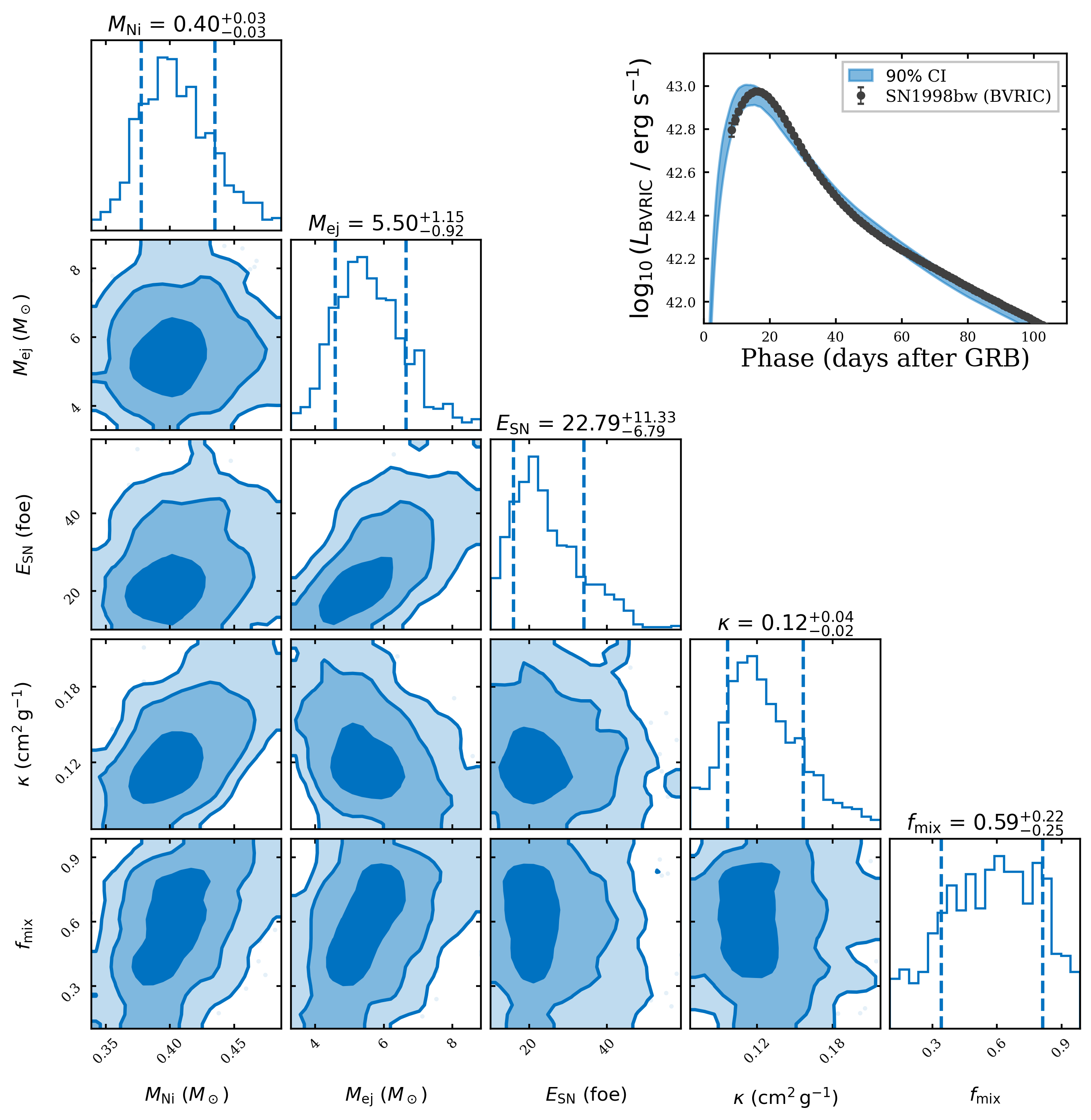}
    \caption{Fit to the BVRIC quasi-bolometric light curve of SN~1998bw using
    \snmix\ with NUTS posterior sampling. The corner plot shows the posterior
    covariance between $\Mni$, $\Mej$, $\Ek$, $\kappa$, and $\fmix$. The
    light-curve panel shows the median model and 90 per cent posterior
    predictive interval. The luminosity scale constrains $\Mni$, while $\Mej$,
    $\Ek$, $\kappa$, $\fmix$, diffusion history, and gamma-ray escape remain
    entangled without velocity and colour information.}
    \label{fig:grb_sn_fit}
\end{figure*}

We next fit mixed light curves with a one-zone model. This is the practical problem of model misspecification~\citep{Nott2024}. A fit can look good while the inferred physical parameters are wrong.

Before fitting mixed light curves, we define a self-consistent one-zone benchmark of the \snmix{} model. In particular, we collapse the leakage calculation into one thermal reservoir, with the same decay constants, positron trapping, gamma-ray convention, velocity convention $v_{\rm ej}=(2\Ek/\Mej)^{1/2}$, and one global optical opacity. We stress that this benchmark is not obtained by taking $\fmix\rightarrow0$ as \snmix\ still evolves separate shell reservoirs and computes their optical and gamma-ray columns regardless of the distribution of the nickel. The one-zone benchmark instead places all deposited energy in one global reservoir. We use this internal one-zone model because it isolates the effect of radial structure while keeping the other conventions matched to \snmix. We still show the \redback\ Arnett implementation in Fig.~\ref{fig:mixing_lc_family} and discuss the bias relative to this model due to its wide use in interpreting light curves. However, this model has its own diffusion normalisation, gamma-ray treatment, and photosphere prescription, and so departs more significantly from the \snmix{} model described here.  

We generate synthetic bolometric \snmix\ light curves from $\fmix=f_{\rm Ni}=0.07$ to $\fmix=1$ and fit each with the internal one-zone benchmark and the \redback{} Arnett model. The kinetic energy is fixed, so changing $\Mej$ changes $v_{\rm ej}=(2\Ek/\Mej)^{1/2}$. Our fits allow the characteristic velocity to change through this relation. If an independent spectroscopic velocity constraint provided a constraint on $\Ek/\Mej$, the model could no longer shorten the diffusion time by changing $\Mej$ and velocity together. More of the mismatch would then have to be absorbed by $\kappa$ or would remain as tension with the light curve. We note that while line velocities can help, they depend on ionisation and abundance structure and may not correspond to a characteristic velocity of the ejecta. We also deliberately use bolometric lightcurves because this isolates the response to missing radial structure without also fitting colour, extinction, temperature evolution, or a SED model. In multi-band fits the same problem should remain, but the bias can move between $\Mej$, $f_{\rm Ni}$, $\kappa$, temperature, extinction, and $\Mni$.

As in other examples of model misspecification in transient light-curve fitting~\citep[e.g.,][]{Sarin2024_cautionary}, the problem is not only whether the fit is bad. A reasonable fit can still lead to the wrong interpretation. A one-zone model has only one diffusion time, so a fast rise is absorbed by low $\Mej$, high $\Ek/\Mej$, low $\kappa$, or high $f_{\rm Ni}$. To explore this impact, we apply two fitting strategies. In strategy~1 we fix $\kappa=0.2\,{\rm cm^2\,g^{-1}}$ (same as input) and fit $f_{\rm Ni}$, $\Mej$, and $v_{\rm ej}$. Fixing the opacity based on the supernova classification is a common approach in light-curve fitting of supernovae. With the second fitting strategy we also fit $\kappa$. In Fig.~\ref{fig:ni_bias}, we summarise our results. In the top left panel we show two representative lightcurves generated with \snmix{} at $\fmix = 0.1$ and $1.0$ and the fits with a one-zone model for both strategies. In the bottom left panel we show the fractional residual between the one-zone fit and the true \snmix{} input. Notably (and as one would ideally expect), there is effectively zero residual at $\fmix = 0.1$, while there is a $10\%$ level residual at $\fmix = 1.0$, although this could be absorbed by the typical measurement uncertainty one would expect on bolometric light curves reconstructed from photometry. On the right hand panels we show the best-fitting one-zone parameter divided by the true synthetic input. Unity marks exact recovery. The one-zone curves match the synthetic luminosities well, so the misspecification is not obvious from fit quality alone. 

At fixed opacity, $\Mni=f_{\rm Ni}\Mej$ is nearly stable, but the fit moves to lower $\Mej$ and higher $f_{\rm Ni}$ as nickel is mixed outward. From the least-mixed shell case to $\fmix=1$, strategy~1 changes from $\Mej^{\rm fit}/\Mej^{\rm input}=0.67$ and $f_{\rm Ni}^{\rm fit}/f_{\rm Ni}^{\rm input}=1.54$ to 0.37 and 2.80, respectively. For $\Mej=5\,\Msun$ and $f_{\rm Ni}=0.07$, the fully mixed light curve would be read as $\simeq1.9\,\Msun$ and $f_{\rm Ni}\simeq0.20$. 
In strategy~2, the mismatch moves partly into opacity: $\kappa^{\rm fit}/\kappa^{\rm input}$ falls from 0.77 to 0.24. The nickel mass is recovered to within $\simeq3\%$ in strategy~1 and $\simeq11\%$ in strategy~2. Allowing $\kappa$ to vary is therefore a way to absorb the bias in other parameters at the expense of a low effective opacity. In multi-band fits, such low effective opacities would also have to remain consistent with the colour and temperature evolution, moving the bias into other parameters, such as the total amount of nickel.

Figure~\ref{fig:ni_bias} uses constant opacity for both the synthetic light curves and the one-zone fits. This deliberately isolates the effect of nickel placement. We repeated the analysis using the temperature-dependent opacity in Equation~(\ref{eq:redback_opacity}) for the synthetic light curves, while retaining a grey one-zone fit. With fixed fitting opacity, $\Mej^{\rm fit}/\Mej^{\rm input}$ changes from 0.49 for the minimum-mixing model to 0.34 for the fully mixed model. When opacity is free, the corresponding mass ratios are 0.97 and 0.49, while $\kappa^{\rm fit}/\kappa^{\rm input}$ changes from 0.34 to 0.43. The fitted parameters therefore depend quantitatively on the opacity prescription broadly we again see that a one-zone fit can reproduce the light curve while assigning unmodelled opacity evolution or radioactive structure to the wrong physical parameters.

Repeating the same exercise with the \redback\ Arnett bolometric model gives the same trend albeit with different normalisations. As described above, this model has a different diffusion normalisation and photosphere treatment, such that even the least-mixed shell case is not expected to be consistent with zero bias. With fixed opacity, the fully mixed case is recovered as $\Mej^{\rm fit}/\Mej^{\rm input}=0.35$ and $f_{\rm Ni}^{\rm fit}/f_{\rm Ni}^{\rm input}=3.29$, i.e. an even stronger low-mass, high-nickel-fraction interpretation than the internal one-zone benchmark. When opacity is free, the same \redback{} Arnett fit instead moves most of the mismatch into opacity, giving $\kappa^{\rm fit}/\kappa^{\rm input}=0.11$, $\Mej^{\rm fit}/\Mej^{\rm input}=1.06$, and $\Mni^{\rm fit}/\Mni^{\rm input}=0.87$. 
Therefore the exact parameter displacement is implementation dependent, but the trend is consistent: one-zone models may fit a mixed radioactive light curve (and in some cases well) but assign the missing radial structure to low ejecta mass, high nickel fraction, or low opacity, which can lead to biased interpretations. 

\subsection{SN~1998bw} \label{sec:sn1998bw}

We now test our model on a real supernova. In particular, we use SN~1998bw/GRB~980425 because it is well observed, associated with a GRB, and has a well-constrained explosion time. We fit the quasi-bolometric light curve from \citet{Clocchiatti2011} with the \snmix\ model, sampling $\Mej$, $\Ek$, $f_{\rm Ni}$, $\kappa$, and $\fmix$ with a 0.10~dex uncertainty added in quadrature to the statistical error on the data. This floor accounts for the systematic uncertainty in constructing a quasi-bolometric light curve and for the limited fidelity of the semi-analytic model. A smaller floor would tighten the reported credible intervals and give greater weight to small model residuals, whereas a larger floor would broaden them. The priors are $\ln(\Mej/\Msun)\sim\mathcal{N}(\ln6.3,0.35^2)$, $\ln(\Ek/{\rm foe})\sim\mathcal{N}(\ln20,0.45^2)$, $\ln(\kappa/0.07\,{\rm cm^2\,g^{-1}})\sim\mathcal{N}(0,0.4^2)$, and a unit-width logistic-normal prior on $f_{\rm Ni}$. The prior on $\fmix$ is uniform between $f_{\rm Ni}$ and 1, enforcing $X_{\rm Ni}\leq1$. The $\Mej$ and $\Ek$ priors are informed by spectral modelling~\citep{Iwamoto1998, Nakamura2001}. Given the \jax{}-implementation we take advantage of a GPU (RTX 5800) and sample using \program{numpyro}~\citep{Phan2019}. 

In Fig.~\ref{fig:grb_sn_fit}, we show a corner plot of our fitted parameters and the lightcurve with a $90\%$ credible interval from our fit. The posterior median gives $\Mni = 0.403^{+0.032}_{-0.026}\,\Msun$, $\Mej = 5.50^{+1.15}_{-0.92}\,\Msun$, $\Ek = 22.8^{+11.3}_{-6.8}\,{\rm foe}$, $\kappa = 0.121^{+0.036}_{-0.025}\,{\rm cm^2\,g^{-1}}$, and $\fmix = 0.59^{+0.22}_{-0.25}$ (Fig.~\ref{fig:grb_sn_fit}). As shown by the analyses above, the luminosity scale constrains $\Mni$ well but the $\Mej$, $\Ek$, $\kappa$, and $\fmix$ constraints are broader because bolometric data mainly constrain the diffusion combination. These constraints are consistent with independent analyses of SN~1998bw~\citep[e.g.,][]{Nakamura2001, Cano2017}, building confidence in our model. We also perform a one-zone maximum likelihood estimate which gives $\Mni=0.416\,\Msun$, $\Mej=4.76\,\Msun$, $\Ek=23.9\,{\rm foe}$, and $\kappa=0.079\,{\rm cm^2\,g^{-1}}$, with a comparable light-curve fit. This analysis demonstrates the level of constraint available from bolometric data alone for the prototypical SN. For the inferred level of mixing in SN~1998bw, the one-zone and \snmix{} interpretations overlap at the posterior level even though their maximum-likelihood points differ. We do not perform the same inference for new Einstein Probe Ic-BL/GRB-SN sample here, but those events are the natural next application: several have fast optical rises and have already motivated engine-based interpretations, while their degree of radioactive mixing has not yet been explored in the same framework.

\begin{figure*}
    \centering
    \includegraphics[width=1.00\textwidth]{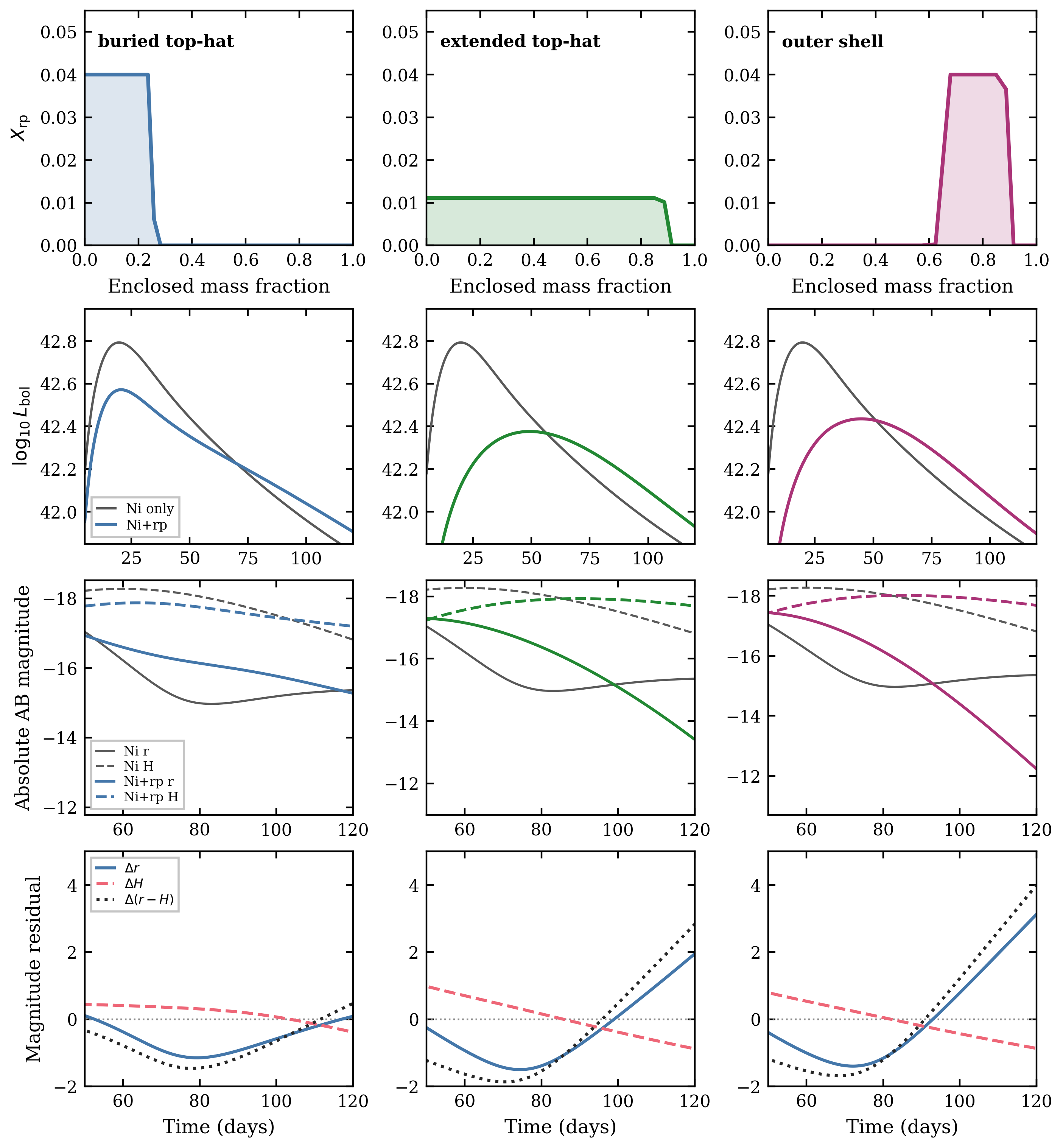}
    \caption{Observational signatures of neutron-rich ejecta at fixed $\Mej=5\,\Msun$, $\Ek=10\,{\rm foe}$, $\Mni=0.35\,\Msun$, $M_{\rm rp}=0.05\,\Msun$, and $\fmix=0.5$. Different columns, from left to right show a central top-hat, an extended top-hat, and an outer shell r-process distribution. The rows beneath show different observational signatures for the corresponding profile. The total r-process mass is fixed in all cases but $X_{\rm rp}$ changes because the same mass is spread over different mass intervals. The final row shows $\Delta r=r_{\rm rp}-r_{\rm Ni}$, $\Delta H=H_{\rm rp}-H_{\rm Ni}$, and $\Delta(r-H)=(r-H)_{\rm rp}-(r-H)_{\rm Ni}$. The same r-process mass can suppress optical light, delay the NIR, or produce a late relative excess depending on placement.}
    \label{fig:rprocess_placement_timing}
\end{figure*}

\begin{figure*}
    \centering
    \includegraphics[width=0.99\textwidth]{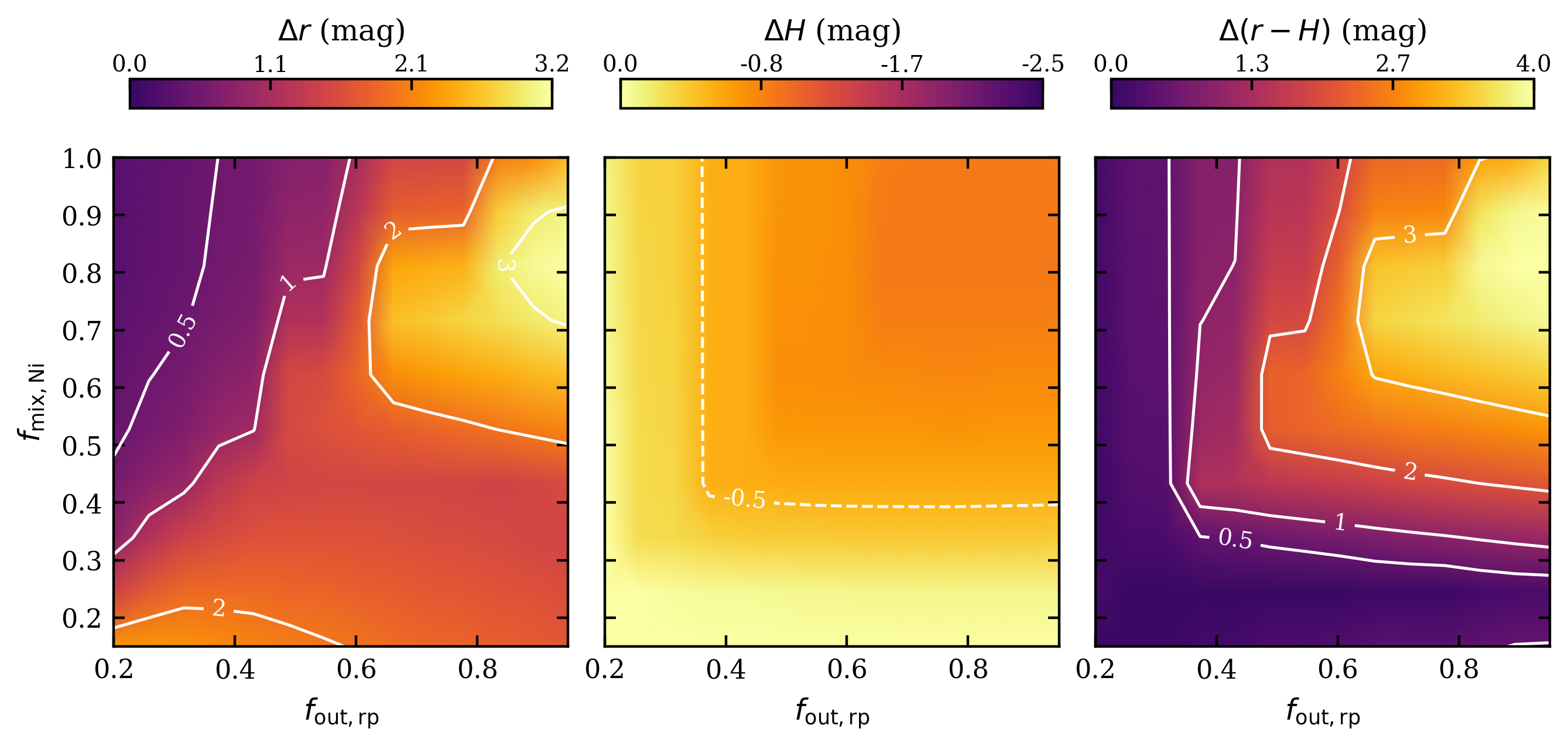}
    \caption{Observational signatures of neutron-rich ejecta as a function of the degree of nickel mixing and the placement of the neutron-rich ejecta. The r-process material is a central top-hat with fixed $M_{\rm rp}=0.05\,\Msun$. The horizontal axis gives its outer enclosed-mass coordinate; the vertical axis gives $\fmix$. Each cell is compared to a Ni-only model with the same $\fmix$. Panels show the maximum $\Delta r$, minimum $\Delta H$, and maximum $\Delta(r-H)$ over 30-120 days post explosion. \label{fig:rprocess_geometry}}
\end{figure*}
\section{R-process Enrichment in Collapsars} \label{sec:rprocess}
We now turn our attention to another source of radioactivity where the placement can make a dramatic difference to observables: r-process elements in collapsars. Numerical simulations suggest that collapsar accretion disks may have sufficient neutron-rich material to be a viable production site for r-process elements~\citep{Siegel2019}. Semi-analytic modelling has previously shown that the distribution of this material into the ejecta can make a significant impact on the observable signature: a near-infrared excess due to the high-opacity r-process material~\citep{Barnes2022}. Here, we use \snmix{} to explore the impact of r-process nucleosynthesis in our model and consider different geometrical configurations (spherical and viewing-angle dependent), alongside different treatments of opacities, heating rates, and thermalisation efficiencies when combined with effects such as nickel mixing. 
As we have established above, nickel mixing can strongly influence the SN lightcurve and the observable signatures of collapsar r-process must be interpreted relative to the SN background, making the exploration of observables as a function of mixing pertinent.  We note that we are not exploring here what drives the particular placement of the neutron-rich ejecta, i.e., the hydrodynamical origin of the distribution (or the origin of the neutron-rich ejecta itself), but simply how the observable signatures change for different configurations. Numerical studies of accretion disk outflows suggest the mixing of r-process material may be strongly correlated with the disk-wind mass and duration and by extension the SN explosion energy~\citep{Barnes2023}, which we also ignore. 

Below, we describe the extensions to the basic \snmix{} model introduced in Sec.~\ref{sec:model} to capture the effects of neutron-rich ejecta. We then present our results exploring the observable signatures for different assumptions. 



\subsection{Extension to \snmix{} for neutron-rich ejecta} \label{sec:rprocess_model}

The r-process calculations use the shell transport from Section~\ref{sec:model} and add three ingredients: placement, r-process heating/opacity, and the optically-thin SED prescription used for broadband photometry. The latter closely follows the treatment in~\citet{Barnes2022}.

For spherical calculations we use two r-process placements. The first is a mass-coordinate top-hat, which enables more direct comparisons to~\citet{Barnes2022} with a slow r-process-rich component underneath a faster supernova envelope. The second is a mass shell,
\begin{equation}
    X_{\rm rp}(m) =
    \begin{cases}
        M_{\rm rp} / [(m_{\rm out}-m_{\rm in}) \Mej] & m_{\rm in}\Mej < m \leq m_{\rm out}\Mej, \\
        0 & \mathrm{otherwise},
    \end{cases}
\end{equation}
which approximates disk-wind material mixed into intermediate or outer ejecta layers in a spherical average. The shell width must exceed $M_{\rm rp}/\Mej$ so that $X_{\rm rp}\leq1$. In the two-component calculations below, ``inner disk wind'' means $m_{\rm in}=0$ and $m_{\rm out}=0.5$, while ``outer disk wind'' means $m_{\rm in}=0.5$ and $m_{\rm out}=1$. These coordinates refer to enclosed mass within the equatorial component. Each shell is assigned an electron fraction, $Y_e$, which sets both the r-process heating-rate interpolation and the lanthanide-rich opacity. In the main r-process calculations we use a linear velocity-space gradient between an inner neutron-rich value and an outer higher-$Y_e$ value. We also implement uniform and two-component $Y_e$ profiles for diagnostic tests.

For shells containing r-process material, we use the analytic heating-rate approximant of \citet{Rosswog2024}, calibrated against nuclear network calculations across $Y_e \in [0.05, 0.50]$ and $v \in [0.05, 0.50]\,c$:
\begin{equation} \label{eq:rp_heating}
\begin{split}
    \dot{q}_{\rm rp}(t, v, Y_e) = q_0 &\!\left(\frac{1}{2} - \frac{\arctan[(t-t_0)/\sigma]}{\pi}\right)^{\!\alpha} \!\!\left(\frac{1}{2} + \frac{\arctan[(t-t_1)/\sigma_1]}{\pi}\right)^{\!\alpha_1} \\
    &+ C_1 e^{-t/\tau_1} + C_2 e^{-t/\tau_2} + C_3 e^{-t/\tau_3},
\end{split}
\end{equation}
where $t$ is in seconds and all thirteen coefficients depend on $(v,Y_e)$ through bilinear interpolation~\citep{Rosswog2024, Sarin2024_cautionary}. At late times this approaches $\dot{q}_{\rm rp}\propto t^{-1.3}$~\citep{Metzger2010, Korobkin2012}. 

The opacity is also set by electron fraction through the lanthanide abundance. We interpolate the grey \citet{Tanaka2020} values over $0.10\leq Y_e\leq0.50$, from $\kappa_{\rm rp}\simeq35$ to $0.5\,{\rm cm^2\,g^{-1}}$. In mixed nickel and r-process shells we dilute this opacity by the local r-process mass fraction,
\begin{equation}\label{eq:kappa_rp}
    \kappa_i = \kappa_{\rm SN} + \min(X_{{\rm rp},i}/0.1, 1)\,[\kappa_{\rm Tanaka}(Y_{e,i})-\kappa_{\rm SN}].
\end{equation}
The value $X_{\rm rp}=0.1$ is not a physical boundary, but an effective saturation scale. This reflects the expectation that a small fraction of lanthanide-rich material is sufficient for line opacity to dominate over the normal SN opacity~\mbox{\citep[e.g.][]{Kasen2013,Li2019,Tanaka2020}}. We have checked that varying this saturation scale does not change the qualitative trends discussed below.

For r-process thermalisation we calculate separate deposition efficiencies for beta particles, alpha particles, fission fragments, and gamma rays, following the density-scale arguments of \citet{Hotokezaka2020}. Heating is computed shell by shell, but the charged-particle thermalisation scale uses the mass-weighted density of the enriched region, $\rho_{\rm rp,m}(t)t^3$, rather than the numerical shell mass. We use effective opacities $\kappa_{\beta,{\rm eff}}=4.5\,{\rm cm^2\,g^{-1}}$, $\kappa_{\alpha,{\rm eff}}=3\,{\rm cm^2\,g^{-1}}$, and $\kappa_{{\rm fiss},{\rm eff}}=10\,{\rm cm^2\,g^{-1}}$ following~\citet{Hotokezaka2020}. The r-process gamma-ray channel is deposited with
\begin{equation}
    f^{\gamma,{\rm rp}}_{{\rm dep},i} = 1 - \exp[-\kappa_{\gamma,{\rm rp}}\Sigma_i(t)],
\end{equation}
with $\kappa_{\gamma,{\rm rp}}=0.07\,{\rm cm^2\,g^{-1}}$. These particle-dependent efficiencies are combined into the effective thermalisation factor $f_{{\rm th},i}^{\rm rp}$ in the deposited-heating equation above.

To compute photometry we follow~\citet{Barnes2022}. Optically thick shells keep the luminosity-conserving photospheric SED from Section~\ref{sec:sed}. As $\tau$ drops through $\simeq2/3$, r-process-associated luminosity is smoothly mapped to the scaled $T_{\rm rp}=2500\,{\rm K}$ blackbody proxy used by~\citet{Barnes2022}, while r-process-free luminosity is mapped to an empirical SN-like nebular SED. Nebular-phase radiative transfer is needed to test how reliable this approximation is for individual events.

To capture the aspherical geometry expected in collapsars, we use a two-component extension: a polar SN component carrying the nickel-powered emission and an equatorial disk-wind component representing neutron-rich ejecta. The two components are evolved independently and then combined with a viewing-angle weight,
\begin{equation}
  L_\nu^{\rm obs}(\theta_{\rm obs}) =
  w_{\rm p}(\theta_{\rm obs}) L_\nu^{\rm p}
  + w_{\rm eq}(\theta_{\rm obs}) L_\nu^{\rm eq},
\end{equation}
where
\begin{equation}
  w_{\rm p}(\theta_{\rm obs}) =
  \exp\left[-\frac{1}{2}
  \left(\frac{\theta_{\rm obs}}{\theta_{\rm j}}\right)^2\right],
  \qquad
  w_{\rm eq}(\theta_{\rm obs}) = 1-w_{\rm p}(\theta_{\rm obs}).
\end{equation}
We use $\theta_{\rm j}=20^\circ$ in the viewing-angle figures below. This angle is the scale over which the polar component dominates the observed flux, not a sharp physical opening angle of the r-process wind. This is a simple prescription. It neglects radiative coupling, occultation, and lateral transport between the polar and equatorial components, but is sufficient for testing how the r-process residual changes from on-axis to equatorial viewing angles.

\subsection{Neutron-rich ejecta in a nickel-mixed SN} \label{sec:rp_geometry}
We first explore neutron-rich ejecta in a spherical setup. This is our most direct comparison to the setup of \citet{Barnes2022}, modulo our different numerical treatment of neutron-rich ejecta and the supernova itself. In Fig.~\ref{fig:rprocess_placement_timing}, we fix $\Mej=5\,\Msun$, $\Ek=10\,{\rm foe}$, $\Mni=0.35\,\Msun$, $M_{\rm rp}=0.05\,\Msun$, and $\fmix=0.5$, and place the same r-process mass as a central top-hat, an extended top-hat, or an outer shell (columns from left to right, respectively). This isolates the timing effect of the r-process distribution. The top panels show the abundance of the neutron-rich ejecta as a function of the enclosed mass fraction, while the three panels below show the bolometric luminosity (with and without the neutron-rich ejecta), absolute AB magnitudes in $r$ and $H$ band, and magnitude residuals $\Delta r$, $\Delta H$, and colour $\Delta(r - H)$ relative to the nickel only model, respectively. 

We find that the r-process signature is not limited to a late NIR excess. Buried r-process material first acts through its opacity, suppressing optical light relative to the nickel-only model. The redder optically thin emission appears only once the relevant enriched layers become transparent. Moving the same $M_{\rm rp}$ outward shifts this transition earlier, but also spreads the material over more mass and reduces the local $X_{\rm rp}$. The absolute amplitudes are more model dependent because they depend on the effective opacity and on the approximate optically thin SED prescription. The timing trends with r-process placement are a more robust feature of our results and are consistent with the calculations in~\citet{Barnes2022}.

In Fig.~\ref*{fig:rprocess_geometry}, we show how the direct observational signatures, i.e., the magnitude residuals, vary with the outer mass coordinate of a central r-process top-hat and the nickel mixing fraction. Each model is compared to a nickel-only calculation with the same $\fmix$, and the plotted quantities are extrema over 30--120 d. The direct band residuals, $\Delta r$ and $\Delta H$, separate optical suppression from an H-band residual, while $\Delta(r-H)$ gives the observable colour diagnostic. Nickel mixing changes both the background luminosity and the colour contrast. In these spherical models the largest $\Delta(r-H)$ occurs when nickel is strongly mixed and the r-process material extends to a large enclosed mass coordinate, suggesting that fast-rising Ic-BL/GRB-SNe are valuable targets for colour-based searches if neutron-rich material is also mixed outward. The direct band residuals should still be used to interpret whether the colour signal is driven by optical suppression, H-band excess, or both, although in real data these residuals require a model for the nickel-only SN background.

\subsection{Dependencies on viewing angle} \label{sec:viewing_angle}

We next relax spherical symmetry. Collapsar disk winds are not expected to be spherical, so the r-process signal should depend on viewing angle as well as mass and radial placement. The strength of this effect should depend on how strongly the disk-wind material mixes with the supernova ejecta~\citep{Barnes2023}. We use the two-component weighting defined above, with $\theta_{\rm j}=20^\circ$, to combine a polar nickel-powered SN component with a neutron-rich equatorial disk-wind component.

\begin{figure}
    \centering
    \includegraphics[width=\columnwidth]{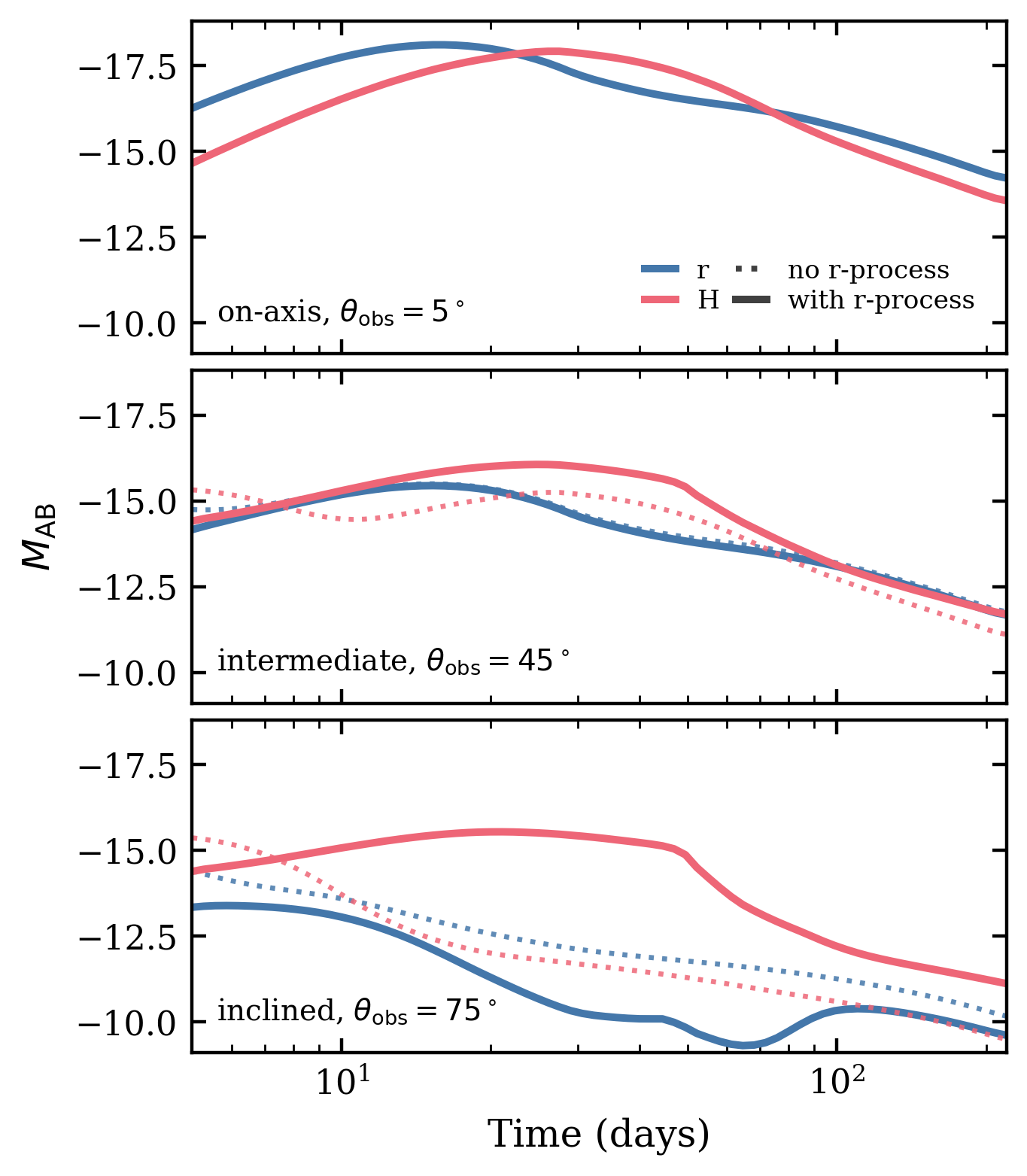}
    \caption{Viewing-angle light curves for the two-component disk-wind collapsar model using the weighting in Section~\ref{sec:rprocess_model} with $\theta_{\rm j}=20^\circ$. The polar/SN component is fixed; the equatorial component is compared with and without $M_{\rm rp}=0.05\,\Msun$ in the outer disk wind. Solid curves include r-process material; dotted curves are the no-r-process control. On-axis views are polar-component dominated, while inclined views show stronger optical suppression and delayed H-band emission.}
    \label{fig:rprocess_viewing_angle_lcs}
\end{figure}

The light curves in Fig.~\ref{fig:rprocess_viewing_angle_lcs} show the consequence for a fixed outer disk-wind configuration. On-axis, the r-process and no-r-process curves are almost identical because the polar SN component dominates the observed flux. At larger inclinations, the disk wind contributes more strongly, producing optical suppression and a delayed H-band residual. This is an important observational result; if the neutron-rich material is confined to an equatorial wind, a classical on-axis GRB-SN is not necessarily the best place to search for the r-process signature.

\begin{figure*}
    \centering
    \includegraphics[width=1.0\textwidth]{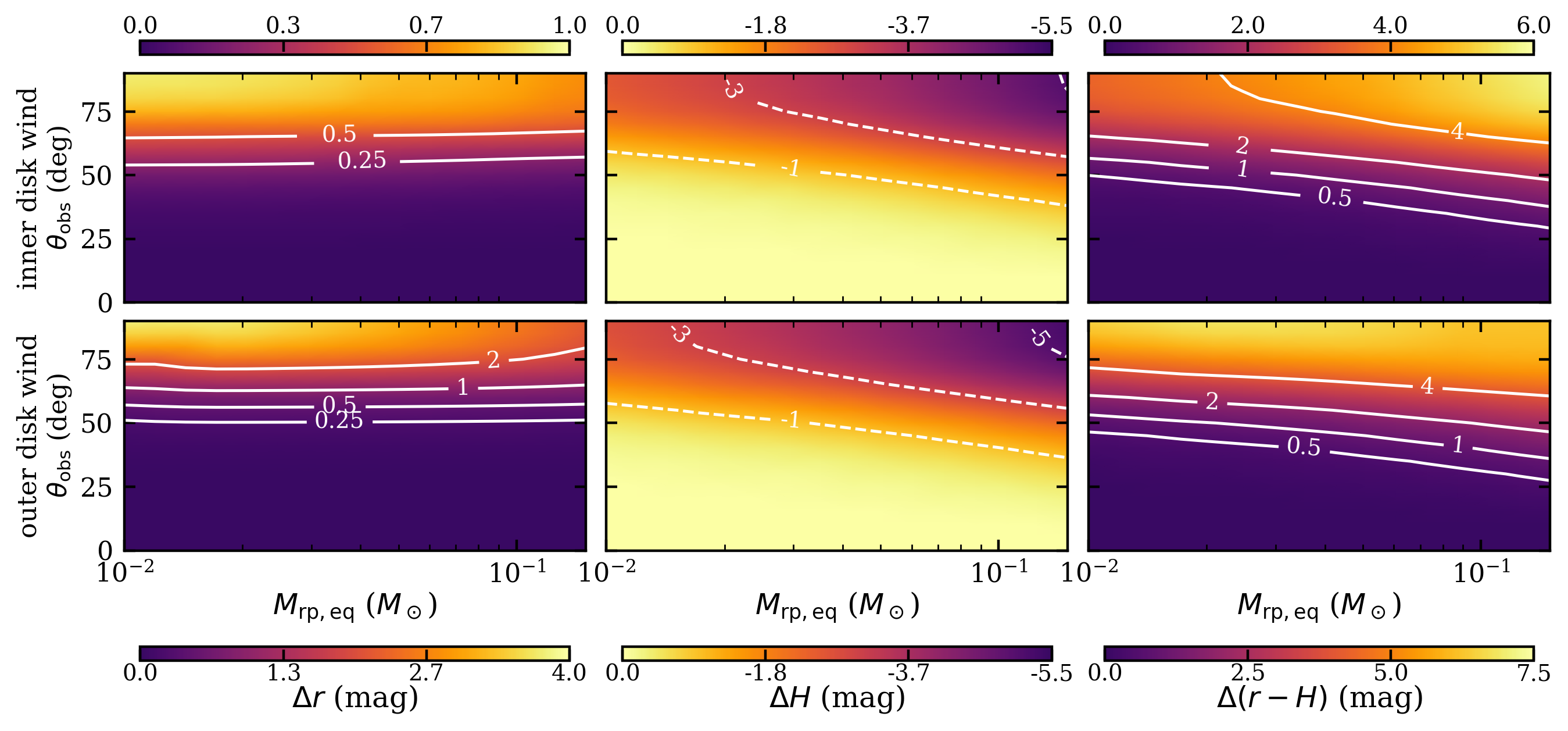}
    \caption{Viewing-angle detectability of an equatorial r-process disk wind. The horizontal axis is the physical r-process mass in the equatorial component and the vertical axis is observer angle from the polar axis. Rows compare the radial placement of the neutron-rich material within the equatorial ejecta. The top row shows an inner disk wind with neutron-rich ejecta placed in the inner half of the equatorial mass, while outer disk wind places it in the outer half (bottom row). Columns show maximum $\Delta r$, minimum $\Delta H$, and maximum $\Delta(r-H)$ over 30-200 d relative to a no-r-process baseline. The signal is weak for on-axis observers because the polar SN component dominates the flux, and strengthens toward equatorial viewing where the disk-wind component contributes more strongly.}
    \label{fig:detectability_viewing_angle}
\end{figure*}

\begin{figure*}
    \centering
    \includegraphics[width=1.0\textwidth]{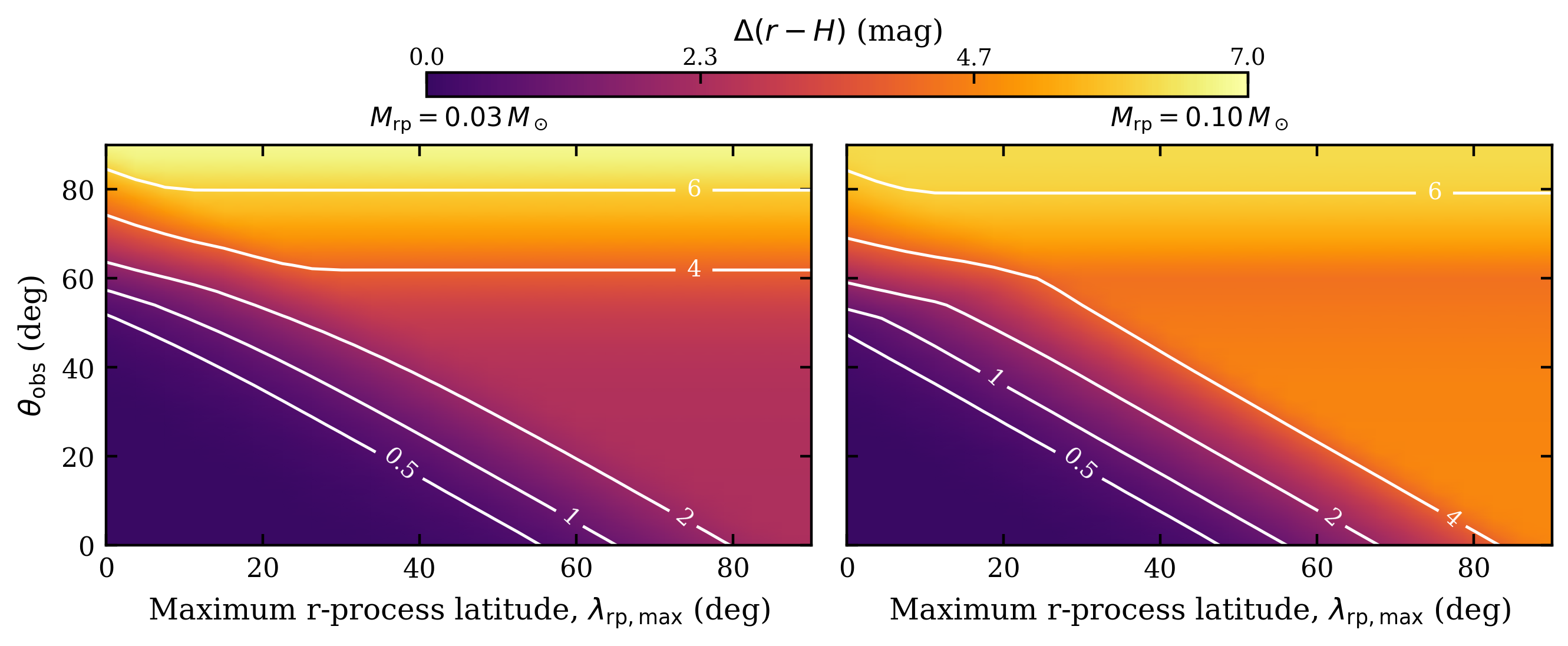}
    \caption{Effect of latitudinal mixing on the observable r-process colour residual. The polar/equatorial background weighting is the same as in Fig.~\ref{fig:rprocess_viewing_angle_lcs}; the r-process mass, radial placement, and ejecta parameters are fixed. The horizontal axis varies the maximum latitude reached by neutron-rich material, $\lambda_{\rm rp,max}$, measured from the equatorial plane, while the vertical axis gives observer angle from the polar axis. The colour scale shows the maximum $\Delta(r-H)$ over 30--200 d relative to the no-r-process baseline. Equatorially confined material is most visible to inclined observers, while material mixed to higher latitudes produces a larger on-axis residual.}
    \label{fig:rprocess_latitude_heatmap}
\end{figure*}

Figure~\ref{fig:detectability_viewing_angle} shows the same observables as Fig.~\ref{fig:rprocess_geometry}, now as a function of $M_{\rm rp}$ and viewing angle. The nickel-only baseline and r-process models have the same polar component and the same equatorial mass, kinetic energy, nickel content, and
baseline SN opacity; only the neutron-rich mass and the associated composition-dependent effective opacity change across the grid. The two rows compare where the neutron-rich material sits within the equatorial component: an inner disk wind with the r-process mass in the inner half of the equatorial ejecta, and an outer disk wind with the same mass placed in the outer half. 
The observable values are extrema over 30--200 days. The direct band residuals show whether a red colour is caused by missing optical flux, an H-band residual, or both. In this two-component approximation, viewing angle is comparable in importance to $M_{\rm rp}$, because the same disk-wind mass can be nearly invisible on-axis and much clearer closer to the equatorial plane.

\subsection{Latitudinal mixing of neutron-rich ejecta} \label{sec:mixing_rp_detect}

A key geometric variable is how far neutron-rich disk-wind material reaches above the equatorial plane. We denote the maximum latitude measured from the equatorial plane by $\lambda_{\rm rp,max}$, so $\lambda_{\rm rp,max}=0^\circ$ is an equatorially confined layer and $\lambda_{\rm rp,max}=90^\circ$ reaches the polar axis. We explore this in Figure~\ref{fig:rprocess_latitude_heatmap} and keep the same polar/equatorial background weighting as above, but vary the angular visibility of the r-process residual. A tight equatorial wind is easiest to see off-axis. If the same material reaches higher latitudes, the on-axis $\Delta(r-H)$ residual grows as polar observers begin to see the neutron-rich component directly. Thus $\theta_{\rm j}$ controls how the polar SN and equatorial wind backgrounds are combined, while the latitude coordinate in Fig.~\ref{fig:rprocess_latitude_heatmap} controls how far the r-process material itself is mixed away from the equator. A non-detection at these phases can therefore mean either that the r-process material is tightly equatorial or that it is buried beneath the SN background. If this vertical mixing is high, a much stronger constraint can be placed on the presence of neutron-rich ejecta. Two-dimensional hydrodynamical simulations of disk-wind outflows are necessary to see how much vertical mixing one could expect in real systems. 

Taken together, Figs.~\ref{fig:rprocess_placement_timing}-\ref{fig:rprocess_latitude_heatmap} show why a collapsar r-process search should not be framed only as a late-time NIR-excess search. The observable colour, especially $\Delta(r-H)$, is valuable, but it folds together two physical effects: optical suppression and H-band brightening. The direct residuals are therefore useful for interpretation, while the colour is the cleaner observing diagnostic. In real events the nickel-only baseline is not known exactly, so the colour evolution should be interpreted together with direct optical and NIR light-curve evolution. Practically, this means paired optical and NIR monitoring from roughly 10-60 d, with enough cadence to identify when the colour separation turns on. Dust, CSM interaction, light echoes, and afterglow contamination can also redden the SED, so spectra and radio/X-ray constraints remain important. Spectra help because lanthanide-rich material, dust, CSM interaction, and ordinary SN cooling should not produce the same line blanketing, continuum shape, or velocity evolution. Radio and X-ray data help identify afterglow or CSM-powered components that can redden the broadband SED without requiring r-process opacity.
\section{Discussion} \label{sec:discussion}
\subsection{Interpretations of SNe lightcurves} \label{sec:discuss_mni}

Our first main result is that nickel mixing can produce faster and brighter rises than one-zone models. This has two immediate implications. First, arguments for the need of an extra source of energy such as an engine or shock emission based on the difficulty of fitting the early time lightcurve may instead be resolved by mixing. Second, one-zone models can produce biased interpretations (inferred parameters) but yield good fits to the lightcurve. These biases (discussed in Section~\ref{sec:ni_bias}) matter because one-zone fits are routinely used to build SESN and Ic-BL samples. Many analyses report $\Mni=0.1$--$0.5\,\Msun$ with fixed or weakly constrained opacity~\citep{Drout2011, Taddia2019}. Our results suggest a hierarchy in what can be trusted from light curves. The luminosity scale, and therefore $\Mni$, is comparatively robust (when fitting to bolometric light curves) but $\Mej$, $f_{\rm Ni}$, $\kappa$, and velocity are not. This bias grows as a function of $\fmix$ and can be well absorbed by posterior uncertainty for $\fmix \leq 0.4$. These biased inferences can have a dramatic impact on interpretations of the explosion energy, progenitor mass, and remnant-formation arguments. 

Mixing also changes the rise and the tail in different directions. Outward nickel reduces the optical diffusion column and can produce a fast, luminous peak, but it also reduces the gamma-ray trapping column and can steepen the radioactive tail. This behaviour is difficult to represent with one global diffusion and leakage scale. The combination of an early fast rise and a late-time decline is therefore more informative than either phase alone. An engine can mimic the early peak, while late bolometric coverage tests whether the event still follows the $^{56}$Co power budget and the expected gamma-ray leakage history.

The SN~1998bw fit illustrates the observational interpretation in practice. The bolometric data constrain the nickel mass, but do not uniquely decompose the ejecta into mass, energy, opacity, mixing, and gamma-ray leakage. Breaking the degeneracy requires information that responds differently to the same parameters, such as photospheric velocities, colour evolution, spectra, and the late-time bolometric decline. For SN~1998bw we infer a moderate $\fmix=0.59^{+0.22}_{-0.25}$. At this level of mixing, the one-zone maximum-likelihood estimate is offset from the \snmix{} posterior median by $\simeq13$ per cent in $\Mej$ and $\simeq5$ per cent in $\Ek$, while $\Mni$ changes by only $\simeq3$ per cent. The full \snmix{} posterior still overlaps the one-zone estimate, so this is not a dramatic failure for SN~1998bw with a bolometric light curve fit. Rather, it illustrates how bolometric data alone allow the missing structure to move between $\Mej$, $\Ek$, $\kappa$, and $\fmix$. The risk is larger for faster and more strongly mixed events, where the same degeneracy can be interpreted as unusually low ejecta mass, low opacity, or additional power. A timely example is EP260321A/SN~2026gsf, for which recent analyses find evidence for strong outward mixing in a fast Ic-BL/GRB-SN-like event~\citep{Martin-Carrillo2026,Rastinejad+26,Chen2026, Yuan2026}. Exploring mixing will become even more important as Einstein Probe and high-cadence optical
surveys help build larger samples of fast Ic-BL/GRB-SNe. A high-energy transient may indicate a jet or engine, but the optical SN still has to be tested separately. A fast optical rise alone does not show that sustained non-radioactive power is required and a one-zone model may lead to an incorrect interpretation. 

\begin{table*}
    \centering
    \caption{Interpretive guide for degeneracies introduced by radioactive structure. The entries are not unique diagnostics; they indicate cases where one-zone interpretations should be checked against velocities, colours, spectra, late-time data, or viewing geometry.}
    \label{tab:observer_summary}
    \small
    \begin{tabular}{p{0.20\textwidth}p{0.24\textwidth}p{0.26\textwidth}p{0.20\textwidth}}
        \toprule
        Observation & Common interpretation & Radioactive-structure alternative & Useful checks \\
        \midrule
        Fast rise and bright peak & Low $\Mej$, low $\kappa$, high $f_{\rm Ni}$, or sustained engine power & Larger true $\Mej$ with $^{56}$Ni mixed to a low overlying column & $v_{\rm phot}$, colour evolution, spectra, late bolometric decline\\
        Good one-zone fit with low $\Mej$ or $\kappa$ & Low-mass or unusually transparent ejecta & Missing outward mixing absorbed into effective diffusion parameters & Fit with opacity free/fixed, compare with velocities and multi-band SED \\
        Peak reproduced but late decline mismatched & Wrong $\Mni$, missing engine power, or simple gamma-ray leakage & Optical diffusion and gamma-ray deposition probe different shell columns & Late bolometric slope, $^{56}$Co budget, nebular spectra \\
        Red optical--NIR colour & Dust, CSM interaction, afterglow contamination, or ordinary SN colour evolution & Lanthanide-rich material can suppress optical light, add delayed H-band emission, or both & Time-dependent $r-H$, band-by-band evolution, spectra, radio/X-ray \\
        Weak r-process signature in an on-axis GRB-SN & Low $M_{\rm rp}$ or no neutron-rich ejecta & Equatorially confined r-process material hidden by the polar SN component & Off-axis events, late optical/NIR coverage, constraints on wind latitude \\
        \bottomrule
    \end{tabular}
\end{table*}

\subsection{Observing signatures of neutron-rich ejecta in collapsars} \label{sec:obs_strategy}

Our modelling suggests that the r-process signature in collapsar SNe is not a one-parameter NIR-excess prediction. The same $M_{\rm rp}$ can produce optical suppression, a delayed H-band residual, or a colour change depending on where the neutron-rich material sits relative to the nickel-powered background. $\Delta(r-H)$ is therefore useful observationally, but it is not sufficient physically because the same colour residual can be driven by missing optical flux, extra H-band flux, or both. Paired optical and NIR monitoring from roughly 10-60 d is therefore more informative than late H-band points alone. This is especially important because the nickel-only baseline that normalizes these residuals is not known for any individual SN. 

Target selection should depend on both the nickel background and geometry. In the spherical calculations, large $\Delta(r-H)$ favours outward r-process placement and can be enhanced when nickel is also strongly mixed. Fast-rising Ic-BL/GRB-SNe are therefore valuable colour-search targets if neutron-rich material is mixed outward. In the two-component disk-wind calculations, however, angular distribution becomes the limiting factor. If the neutron-rich wind is equatorial, off-axis GRB-SNe and Ic-BL SNe without bright afterglows are better targets than classical on-axis GRB-SNe. If the wind reaches high latitudes, on-axis GRB-SNe become more constraining. Thus an on-axis non-detection constrains high-latitude r-process material, but only weakly constrains a strongly equatorial wind.

Our result is consistent with the broad picture of \citet{Barnes2022}, where collapsar r-process material can leave an observable late-time signature, but it changes the interpretation of such a detection. \citet{Barnes2022} use a spherical setup with a fixed supernova background and a single gray opacity for the enriched layer. In \snmix{}, the nickel background, the effective opacity, and the geometry all change the observable. The SN background matters because nickel mixing changes the contrast against which a red component is measured. In our model, the relevant opacity is an effective opacity because mixed spherical shells dilute the intrinsic lanthanide-rich opacity by $X_{\rm rp}$ through Equation~(\ref{eq:kappa_rp}). The geometry matters because an equatorial disk wind can be nearly invisible on-axis, while latitudinal mixing makes the same r-process mass visible to polar observers.

The remaining uncertainty is where the neutron-rich material actually goes. Two-dimensional hydrodynamical simulations can independently probe what level of vertical mixing is realistic for disk-wind outflows, while radiation-transfer calculations are needed to replace our optically-thin SED proxy with event-specific spectra and lightcurves. Our model suggests that we should not expect every collapsar r-process event to show a clean NIR excess. Time-dependent optical suppression, H-band residuals, and $r-H$ colour evolution need to be interpreted together against a structured SN background.

Both parts of the paper point to the same issue. Radioactive structure creates light-curve degeneracies, and those degeneracies are broken only by combining observables. For nickel mixing, the powerful observables are photospheric velocities, colour evolution, and the late bolometric decline, which separate a short diffusion column from low ejecta mass or sustained engine power by comparing the early peak and late decline to the radioactive heating budget. For neutron-rich ejecta in collapsars, the strong observables are paired optical and NIR photometry, direct band residuals, and spectra that can distinguish lanthanide opacity from dust, CSM interaction, afterglow contamination, or ordinary SN colour evolution. Our model is therefore best used to identify which combinations of observations are diagnostic, explore large parameter space efficiently, and infer properties from light curves and compare actual observables, rather than as a replacement for radiation-transfer modelling. In Table~\ref{tab:observer_summary}, we  summarize the main interpretation traps, the alternatives as suggested by our model, and the observations that can test them.

\subsection{Limitations and Assumptions} \label{sec:limitations}

The model makes a number of simplifying assumptions. Throughout, we assume wavelength-independent gray opacities within composition zones, spherical or simplified two-component geometry, and homologous expansion, with limited shell-to-shell interaction. We omit any non-thermal ionisation and recombination effects. These choices make the model fast enough for parameter studies, but limit the accuracy of absolute band predictions.

The nickel profile is also simplified. The baseline top-hat separates total nickel mass from radial extent and makes the bias experiment clean. The core-plus-tail curve in Fig.~\ref{fig:mixing_lc_family} checks that the result is not only a discontinuous-edge effect. Real explosions will have smoother gradients, plumes, and asymmetric nickel fingers. The fitted $\fmix$ should therefore be read as an effective radial extent of radioactive heating, not a reconstructed abundance boundary. Profile shape, opacity evolution, and density structure can trade off in detailed fits.

The transport is a leakage scheme: each shell drains on a cumulative-column diffusion timescale without explicit photon exchange with neighbouring shells. This is useful for controlled comparisons, but becomes approximate especially for deeply buried, high-opacity r-process material. A true diffusion calculation could delay and suppress the emerging red component. The two-component model also treats the SN and wind as radiatively decoupled and combines them with a prescribed angular weighting. Lateral transport, occultation, and hydrodynamic mixing would likely soften the viewing-angle and latitude-mixing trends.

The transition SED in the r-process calculation is an observability prescription following~\citet{Barnes2022}, not a real nebular phase spectral synthesis. It assigns simple spectral shapes after a shell becomes transparent. The opacity blending in mixed spherical shells is likewise an effective-opacity prescription, and may not reflect real microphysics. There are also a number of uncertainties associated with the r-process heating and thermalisation, which likely carry $\sim10$-$100\%$ systematic uncertainty~\citep{Barnes2021, Rosswog2024, Sarin2024_cautionary}, and are also present in pure kilonova analyses. These uncertainties are unlikely to change the qualitative trends but can significantly affect the sizes of any colour and magnitude residuals. 

\section{Conclusions} \label{sec:conclusions}
We introduced \snmix{}, a multi-shell semi-analytic model for homologously expanding radioactive ejecta. The model is designed to test how radial heating and opacity structure change light curves and the parameters inferred from them.
We use this model to study how radioactive structure changes light-curve interpretation of nickel-only SNe and to explore the observables of neutron-rich ejecta. Our main conclusions are:
\begin{enumerate}
    \item The radial distribution of $^{56}$Ni changes both the optical diffusion column and the gamma-ray trapping column. At fixed $\Mni$, $\Mej$, and $\Ek$, outward mixing shortens the rise by reducing the overlying mass above the radioactive material, not the total ejecta mass. It can also make the late radioactive tail fainter by increasing gamma-ray escape from high-velocity nickel.

    \item One-zone fits can look good while assigning the light-curve shape to the wrong parameters. In our bolometric experiment, $\Mni$ is comparatively robust, recovered to within $\simeq3\%$ with fixed opacity and $\simeq11\%$ with free opacity. $\Mej$, $f_{\rm Ni}$, and $\kappa$ are not. For the fully mixed case, fixed-opacity fits give $\Mej^{\rm fit}/\Mej^{\rm input}\simeq0.37$ and $f_{\rm Ni}^{\rm fit}/f_{\rm Ni}^{\rm input}\simeq2.8$; free-opacity fits give $\kappa^{\rm fit}/\kappa^{\rm input}\simeq0.24$. Fits with the \redback\ Arnett model show the same qualitative behaviour, with implementation-dependent normalisations. 

    \item A fast rise should not by itself be used to infer sustained magnetar or fallback power. Strong outward nickel mixing can produce a fast optical peak without lowering the total ejecta mass. Late photometry, velocities, colours, spectra, and high-energy/radio data are needed to test whether additional power is required.

    \item SN~1998bw illustrates the observational regime. Bolometric data constrain $\Mni\simeq0.40\,\Msun$, but velocities, colour evolution, and late-time decline are necessary to robustly separate ejecta mass, opacity, mixing, diffusion history, and gamma-ray escape.

    \item R-process enrichment from collapsar disk winds does not always produce a NIR-excess prediction. In spherical setups the clearest signature can be optical suppression, and the largest colour residuals occur when neutron-rich material is mixed outward against a mixed nickel-powered background. In disk-wind models, viewing angle and latitudinal mixing control the optical-NIR contrast. The useful diagnostic is the time evolution of optical flux, H-band residuals, and optical--NIR colour, not only a NIR excess, which can be degenerate with other physical mechanisms.

    \item The r-process signal depends on the nickel-powered background. Nickel mixing changes both the luminosity background and the colour contrast. Strongly mixed, fast-rising events can be good colour-search targets if the neutron-rich material is also radially extended, but the direct band residuals are needed to decide whether the colour comes from optical suppression, H-band emission, or both.

    \item Off-axis GRB-SNe and Ic-BL SNe with optical and NIR coverage from $\sim10$ to 60 d are promising if the neutron-rich wind is equatorial. Classical on-axis GRB-SNe are not necessarily the best NIR-excess targets in that limit, but become constraining if r-process material reaches high latitudes. Detectability is therefore a constraint on both r-process mass and angular distribution, within the assumed two-component angular weighting.
\end{enumerate}

\section*{Acknowledgments}
We thank the anonymous referee for their comments that helped improve the manuscript. We are grateful to Ilya Mandel for early discussions that helped shape this manuscript. We also thank Antonio Martin-Carrillo and Gokul Srinivasaragavan for comments on the manuscript. NS is supported by the Kavli Foundation. The author acknowledges the use of ChatGPT to assist with plotting scripts and review of the text. The author takes full responsibility for the content.
\section*{Data Availability}
The \snmix\ source code will be made publicly available at \url{https://github.com/nikhil-sarin/snmix} after publication. Code is provided in \redback{} (\url{https://github.com/nikhil-sarin/redback}) for the pure nickel mixing model, for simulating data, and sampling on CPUs. All figure scripts are included in the \snmix\ repository. This work also made use of \program{JAX}~\citep{jax2018github}, \program{jax-bandflux}~\citep{leeney2025jax}, \program{NumPyro}~\citep{Phan2019}, \program{Astropy}~\citep{AstropyCollaboration2022}, \program{SciPy}~\citep{2020SciPy-NMeth}, and \program{Matplotlib}~\citep{Hunter2007}.

\bibliographystyle{mnras} 
\bibliography{paper}
\bsp	
\label{lastpage}
\end{document}